\documentclass[a4,11pt]{article}

\usepackage{microtype}
\usepackage{graphicx}
\usepackage{subcaption}
\usepackage[table]{xcolor}
\usepackage{booktabs} 
\usepackage{makecell}
\usepackage{hyperref}
\usepackage{fullpage}

\usepackage{algorithm}
\usepackage{algorithmic}
\usepackage{amsmath}
\usepackage{amssymb}
\usepackage{mathtools}
\usepackage{amsthm}
\usepackage{natbib}
\usepackage[capitalize,noabbrev]{cleveref}
\usepackage{authblk}

\def\preprint{}

\title{Likelihood-free inference of phylogenetic tree posterior distributions}
\author[1,*]{Luc Blassel}
\author[1]{Noémie Sauvage}
\author[1]{Pierre Barrat-Charlaix}
\author[2]{Bastien Boussau}
\author[2]{Nicolas Lartillot}
\author[1,*]{Laurent Jacob}
\affil[1]{Laboratory of Computational, Quantitative and Synthetic Biology, Sorbonne Université, Paris, France}
\affil[2]{Laboratoire de Biométrie et Biologie Évolutive, Université Lyon 1, Villeurbanne, France}
\affil[*]{\small\texttt{luc.blassel@sorbonne-universite.fr, laurent.jacob@cnrs.fr}}

\date{}

\newcommand{\ivec}[2]{\mathbf{#1}_{#2}} 
\newcommand{\relu}{\text{reLU}}
\newcommand{\conv}{\text{Conv2D}}
\newcommand{\lin}{\text{Linear}}
\newcommand{\linn}{\text{LinearNoBias}}
\newcommand{\mean}{\text{mean}}
\newcommand{\sig}{\text{sigmoid}}
\newcommand{\smx}{\text{softmax}}
\newcommand{\smn}{\text{softmin}}
\newcommand{\lnorm}{\text{LayerNorm}}
\newcommand{\ct}{\text{concat}}
\newcommand{\bi}[1]{\emph{\bfseries #1}}
\DeclareMathOperator*{\argmin}{arg\,min}

\begin{document}
    \maketitle
    \begin{abstract}
	Phylogenetic inference, the task of reconstructing how related sequences evolved from common ancestors, is a central objective in evolutionary genomics. The current state-of-the-art methods exploit probabilistic models of sequence evolution along phylogenetic trees, by searching for the tree
	maximizing the likelihood of observed sequences, or by estimating the
	posterior of the tree given the sequences in a Bayesian
	framework. Both approaches typically require to compute likelihoods,
	which is only feasible under simplifying assumptions such as
	independence of the evolution at the different positions of the
	sequence, and even then remains a costly operation. Here we present
	the first likelihood-free inference
	method for posterior distributions over phylogenies. It exploits a novel expressive
	encoding for pairs of sequences, and a parameterized probability distribution
	factorized over a succession of subtree merges. 
    The resulting network provides well-calibrated estimates of the
posterior distribution leading to more accurate tree topologies
than existing methods, even under models
amenable to likelihood computation.
We further show that its edge against likelihood-based methods dramatically increases
under models of sequence evolution with intractable likelihoods.

\end{abstract}

\section{Introduction}
The genomes of living species evolve over time through a process
that involves mutations and selection. Reconstructing the evolutionary
history of a set of contemporaneous sequences is a central task in
genomics \citep{kapli_phylogenetic_2020}: it is used to understand how extant species have evolved from
common ancestors \citep{alvarez-carretero_species-level_2022}, how bacterial resistances to drugs have
emerged and been disseminated \citep{10.1111/j.1574-6968.2007.00757.x}, and how epidemics are spreading
\cite{hadfield_nextstrain_2018}. A key object in this endeavor is the phylogeny, a
bifurcating tree summarizing the succession of transformations of the
sequence that lead to the current observed diversity from a single
ancestor.

Modern phylogenetic reconstruction is now dominated by approaches based on probabilistic models of sequence evolution. Model-based approaches offer a principled framework for inference by making assumptions about evolutionary processes transparent, while also providing methods for model checking, criticism and iterative model elaboration~\citep{Gelman:2004tc}.

Models of sequence evolution are typically
continuous time Markov processes parameterized by the phylogeny and
by the rates at which amino-acids or nucleotides undergo substitutions.
Existing reconstruction methods look for the
phylogeny maximizing the likelihood of the sequences~\citep{minh2020IQ-TREE,price2010FastTree} or, in a Bayesian perspective, aim at sampling from the
posterior distribution of the phylogeny given the sequences through
Monte Carlo strategies~\citep{huelsenbeck2001MRBAYES,hohna2016RevBayes,bouchard2012phylogenetic} or approximate this posterior
distribution through variational inference~\citep{zhang2018variational,koptagel2022vaiphy,zhou2024phylogfn,duan2024phylogen}.

Across all these approaches, a major hurdle is the computational cost
of evaluating the likelihood function, which is required either for maximizing the likelihood, computing acceptance probabilities in sampling strategies, or computing the evidence lower
bound (ELBO) objective in the case of variational inference. For any single phylogeny, the likelihood is computed through a costly
pruning algorithm~\citep{felsenstein1981evolutionary} and many such evaluations are required to heuristically explore the set of possible tree
topologies---$(2N-5)!!$ for a phylogeny over $N$ leaves. Furthermore, just making
this computation feasible has resulted in a focus on probabilistic models that make
simplifying assumptions such as independence and identical
distribution of the evolutionary process at each position in the sequence, or the
absence of natural selection. These simplifications are known to
produce unrealistic sets of sequences~\citep{trost2024simulations}, artifacts in reconstructed phylogenies~\citep{telford2005consideration}, and impede our ability to understand the evolutionary history of living species.

Simulation-based or likelihood-free inference has emerged as a
powerful paradigm for estimation under probabilistic models under
which likelihood evaluations are intractable but sampling is cheap~\citep{cranmer2020frontier,lueckmann2021benchmarking}. This paradigm has leveraged advances in deep learning to approximate posterior distributions by neural networks trained over data simulated under probabilistic
models~\citep{greenberg2019automatic,lueckmann2021benchmarking}.
Among these methods, neural posterior estimation \citep[NPE,][]{lueckmann2021benchmarking} defines a family of distributions parameterized by a neural
network whose weights are then optimized to approximate the posterior. In addition to working around the need for likelihood evaluation, NPE is amortized: training the network can take time but performing inferences with the trained network is typically very fast. By contrast, existing variational inference and Monte Carlo samplers require new computation every time a new inference is done.
On the other hand, NPE require to define a parameterized family of distributions that is appropriate for
the posteriors of phylogenies given sequences, which is not straightforward.

Here we introduce Phyloformer\,2, an NPE for phylogenetic
reconstruction, with the following contributions:
\begin{itemize}
	\item We propose a parameterized family of posterior
	      distributions on phylogenies given a set of sequences, factorized
	      through a succession of pairwise mergings.
	      Optimizing the weights
	      of our network to maximize the log-probability within this family yields a likelihood-free
	      estimate of the corresponding posterior that can then be used for sampling trees.
	      To our knowledge, this is
	      the first likelihood-free posterior estimation method trained end-to-end from sequences to the
	      phylogeny beyond quartets.
	\item To extract the parameters of the approximate
	      posterior distribution from the input sequences, we introduce a
	      novel evoPhyloFormer (evoPF) architecture akin to the EvoFormer module used in Alphafold~2~\citep{jumper2021highly}, that is both more scalable and expressive than the one used in Phyloformer.
	      The
	      overall architecture allows us to scale up to over 200 sequences of
	      length 500 or more than 300 sequences of length 250 on a single V100 GPU with 16Gb of VRAM.
	\item On data generated under a probabilistic model of sequence
	      evolution with tractable likelihood, Phyloformer~2 outperforms
	      both state-of-the-art likelihood-based and likelihood-free
	      reconstruction methods in topological accuracy and produces calibrated
	      estimates of the posterior 
          Because it is likelihood-free, it is the first well-specified estimator 
under models with intractable likelihoods. We show that in this
case, its topological error is 2 to 3 times lower than the one
incurred by---misspecified---likelihood-based estimators.
	\item Because Phyloformer~2 is amortized, once trained, it performs
	      inference 1 to 2 orders of magnitude faster than the---less
	      accurate---state-of-the-art likelihood-based estimators.
\end{itemize}

\subsection*{Related work}

Initial
attempts to phylogenetic NPE were restricted to
quartets, \emph{i.e.}, topologies over four leaves, allowing them to
cast the problem as a classification over the three possible
topologies~\citep{suvorov2019accurate,zou2020deep,tang2024novel}. In this case, a vector embedding of the input sequences
was extracted by a neural network and used to produce three scalar outputs, and the probability of each
topology was simply modeled as a softmax over these three outputs. Since the number of possible topologies
grows super-exponentially with the number of leaves, this strategy cannot be
generalized to larger numbers of sequences and even if it could,
treating all topologies as separate classes would disregard the fact
that some are more similar than others. In addition, \citet{grosshauser2021reevaluating} showed that the method of~\citet{zou2020deep}
underperformed on more difficult tasks with short sequences and long evolution times.

Alternatively, \citet{nesterenko2025phyloformer} proposed Phyloformer, a likelihood-free inference method defined for any number of leaves. Phyloformer doesn't estimate phylogenies, but
evolutionary distances, \emph{i.e.}, sum of
branch lengths on the phylogeny between pairs of leaves. Phylogenies can be reconstructed from such distances with the neighbor joining algorithm \citep[NJ,][]{saitou1987neighbor}
, but the authors observed that this strategy led to a limited topological reconstruction accuracy.
In addition, Phyloformer only allows for point
estimates---namely the median of the posterior distribution of evolutionary distances---rather than the entire
distribution. Finally, the network applied axial self-attention~\citep{ho2019axial}
to all pairs of sequences, which led to a large memory footprint even
using a linear approximation of self-attention~\citep{katharopoulos2020transformers}.

Phyloformer~2 addresses each of these limitations of Phyloformer by estimating the entire posterior distribution of phylogenies---instead of a single estimate of evolutionary distances--- thanks to its BayesNJ module, and by using a novel evoPF encoding that is both more lightweight and more efficient than Phyloformer.


\section{Background and notation}

\subsection{Notation}

Let $x = \{x_1,\ldots,x_N\}$ be a set of $N$ sequences of $L$ letters in some fixed alphabet representing organic compounds (\emph{e.g.}, 20 possible amino acids for proteins, 4 possible nucleotides for DNA). We assume that the sequences are aligned, \emph{i.e.}, correspond to $N$ initial sequences of possibly different lengths whose positions were matched to minimize some score quantifying how similar all sequences are at each position, potentially by introducing gaps denoted by a special character in the alphabet~\citep{kapli_phylogenetic_2020}.

A phylogeny $\theta = (\tau, \ell)$ over $x$ is an unrooted binary tree $\tau$ with $N$ leaves and a set $\ell$ of branch lengths in $\mathbb{R}^{2N-3}_+$. The tree $\tau$ can equivalently be represented by a succession of merges. Intuitively, given $N$ species, one chooses two species, pair them to form a cherry, replace them by a single species representing their ancestor, and proceed recursively with the $N-1$ resulting species until the entire tree has been produced. Formally, we denote this succession of merges as $\left\{m^{(k)}\right\}_{k=1}^{N-3}$ where $m^{(k)}\in\mathcal{C}^{(k)}\stackrel{\Delta}{=}\left\{\left(v^{(k)}_i, v^{(k)}_j\right)\in \mathcal{S}_{(k)}^2|i\neq j\right\}$ is a pair of two distinct elements in $ \mathcal{S}_{(k)}$ the set of ``mergeable'' nodes at the $k^{th}$ merge, \emph{i.e.}, either leaves or internal nodes whose children were already merged, and $\mathcal{C}^{k}$ is the set of candidate merges at step $k$. In order to build $\mathcal{S}_{(k)}$ , we define$\mathcal{S}_{(1)}$ to be the set of leaves in $\tau$ and for $k>1$,  $\mathcal{S}_{(k)} \stackrel{\Delta}{=} \left\{\left\{\mathcal{S}_{(k-1)} \cup u^{(k-1)}\right\}\setminus \left(v^{(k-1)}_i, v^{(k-1)}_j\right)\right\}$, and $u^{(k)}$ is the common neighbor in $\tau$ to the nodes merged in $m^{(k)}$---which is always defined because we start from leaves and recursively replace pairs of elements by their common neighbor in the binary tree.

We further denote $\ell^{(k)} = \left\{\ell^{(k)}_i,\ell^{(k)}_j\right\}\in\mathbb{R}_+^2$ the $k$-th cherry, \emph{i.e.}, the set of two branch lengths connecting $m^{(k)}$ to $u^{(k)}$ . In our context, the $N$ leaves will represent the taxa with sequences in $x$, $u^{(k)}$ is the common ancestor of $m^{(k)}$, and $\ell^{(k)}$ is the set of evolutionary times between this ancestor and its two descendants.

\subsection{Neural posterior estimation}

For a probabilistic model $p(x|\theta)$
of the data $x$ given the parameter $\theta$ and a prior distribution
$p(\theta)$, NPE provides a way to estimate the posterior $p(\theta|x)$ in cases where evaluating $p(x|\theta)$ for a given $(x,\theta)$ is too costly or intractable, but where it is possible to sample from this model. It relies on
a family of distributions $q_\psi(\theta|x)$ whose parameters $\psi$ are provided by a neural network acting on the data $x$---by abuse of notation we denote $\psi(x)$ the parameters output by this neural network for an input $x$. NPE builds its approximation of $p(\theta|x)$ by looking for the $q_\psi$ minimizing the average Kullback-Leibler (KL) divergence with the true posterior:
\begin{displaymath}
	 \mathbb{E}_{p(x)}\left[\text{KL}(q_\psi(\theta|x)||p(\theta|x))\right] = \mathbb{E}_{p(x)}\left[\mathbb{E}_{p(\theta|x)}\left[\log p(\theta|x)-\log q_\psi(\theta|x))\right]\right],
\end{displaymath}
where the average is taken over the marginal $p(x) = \int p(x|\theta)p(\theta)d\theta$. Since the $p(\theta|x)$ term does not depend on $\psi$, minimizing this average KL divergence with respect to $\psi$ is equivalent to maximizing the average approximate log-likelihood $q_\psi$ over the joint distribution $p(x,\theta)=p(x|\theta)p(\theta)$~\citep[see \emph{e.g.},][]{radev2020bayesflow}. This is generally achieved by minimizing a Monte Carlo approximation $\sum_{i=1}^n \log q_{\psi(x_i)}(\theta_i|x_i)$ of this average, built over a
large number of examples $\{(x_i, \theta_i)\}_{i=1}^n$ sampled from $p(x,\theta)$ by successively sampling a $\theta_i$ from the prior and an $x_i$ from the model given $\theta_i$.

Consequently, NPE is guaranteed to converge to the true posterior $p(\theta|x)$, provided that the family $q_\psi$ is expressive enough to represent $p(\theta|x)$ and that the optimization algorithm finds the optimum. More precisely, its approximation error depends both on the expressivity of the chosen family of distributions, \emph{i.e.}, on the average KL divergence $\mathbb{E}_{p(x)}\left[\text{KL}(q_{\psi^*(x)}(\theta|x)||p(\theta|x))\right]$  between the best distribution $q_{\psi^*(x)}$ in the family, and the true posterior for each $x$, and on the expressivity of the neural network, \emph{i.e.}, on its ability to map every $x$ to the corresponding best parameters $\psi^*(x)$.

\section{Methods}

\begin{figure*}[ht]
	\centering
    \ifx\preprint\undefined
    	\includegraphics[width=.7\linewidth]{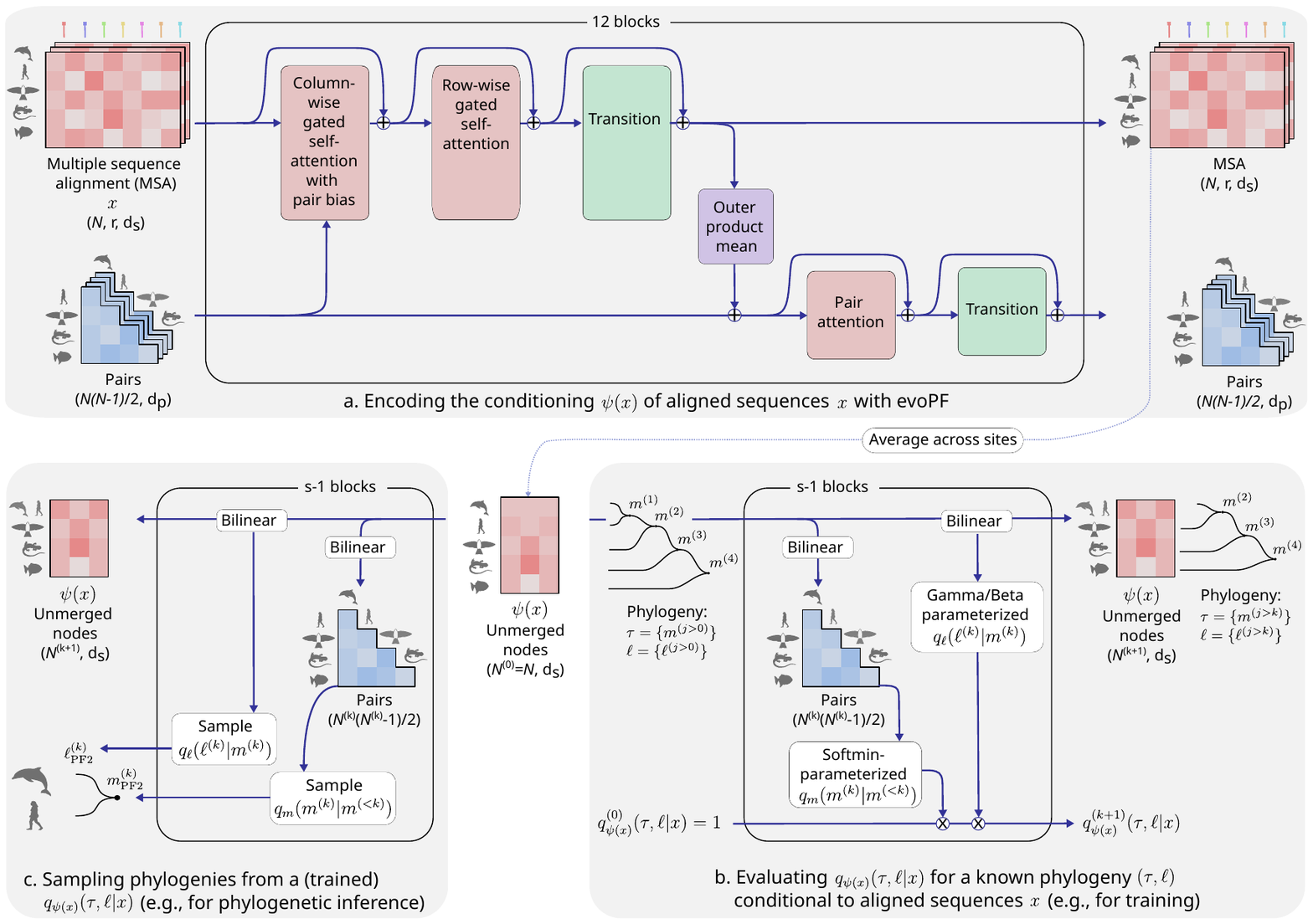}
    \else 
    	\includegraphics[width=.9\linewidth]{figures/evopf.pdf}
    \fi
	\caption{
		Architecture of Phyloformer\,2. \bi{Panel~a}: evoPF, an EvoFormer-inspired module updating $N\times L$ embeddings for a set of aligned sequences (MSA) $x$ and $N(N-1)/2$ embeddings for pairs of sequences. Each of its blocks applies self-attention within both the MSA and representation, and ensures information sharing between them. After 12 blocks, we extract one embedding for each sequence by averaging the MSA embeddings across sites.
		\bi{Panel~b}: BayesNJ (Algorithm~\ref{alg:BayesNJ}) computes the posterior probability of a phylogeny given an MSA represented by the sequence and pair embeddings provided by evoPF. The probability is a product over a recursive operation where two taxa are merged into their parent, and the taxon representation is updated accordingly.
		\bi{Panel~c}: at inference time, we apply the same succession of operations as for evaluating the probability, but either sampling or taking the modes of the distributions (Algorithm~\ref{alg:BayesNJ_sampling}).
	}
	\label{fig:PF2}
\end{figure*}

Phyloformer\,2 combines two novel modules. The first one, evoPF, encodes distribution parameters $\psi(x)$ from a set of aligned related sequences $x$, which is sometimes referred to as a multiple sequence alignment (MSA). The second one, BayesNJ, defines a family of posterior distributions $q_{\psi(x)}\left(\theta=(\tau, \ell)|x\right)$ on phylogenetic trees, parameterized by the output of evoPF. Figure~\ref{fig:PF2}a-b represents the overall architecture.

\subsection{Encoding tree distribution parameters with evoPF}

Our encoder should process a set $x$ of aligned sequences and be expressive enough to capture sufficient information on their evolutionary relationships. \citet{nesterenko2025phyloformer} encoded $x$ with one embedding for each aligned position in each pair of aligned sequences. This approach avoided the need to flatten information across positions, while maintaining a representation at the pair level, consistently with their predicting pairwise evolutionary distances. They noted that tuning down the architecture to one embedding for each position in each sequence---boiling down to the MSA transformer~\citep{rao2021msatransformer}---dramatically improved the method scalability but strongly affected the accuracy of their trained network. Inspired by this trade-off, we introduce evoPF, an encoder that maintains a single embedding per pair, and a separate representation with one embedding per position within each sequence (Figure~\ref{fig:PF2}a).

EvoPF is a transpose version of the EvoFormer module in Alphafold~2~\cite{jumper2021highly}---whose objective was to capture spatial distances between pairs of aligned positions rather than evolutionary distances between pairs of homologous sequences---with a few simplifications. The MSA stack maintains an embedding for each position within each sequence in $x$. It relies on axial attention, alternating one layer of column-wise and one layer of row-wise gated self-attention. The column-wise mechanism takes each sequence separately, and applies self-attention between embeddings of all its positions, allowing information to flow within each sequence. Symmetrically, the row-wise mechanism lets the information flow between sequences by treating each position separately and applying self-attention over the embeddings of all sequences at this position. The two self-attention layers are followed by a transition layer applying a linear function and a ReLU to each embedding in $x$ separately.

Parallel to this MSA stack, we maintain an embedding for each pair of sequences in $x$, updated in each evoPF block by applying self-attention between the embeddings of all pairs, followed by a transition layer like in the MSA stack. By contrast EvoFormer relied on triangular attention where pair $(i,j)$ only attended to pairs $(i,k)$ and $(k,j)$. The MSA stack affects the pair stack through the addition of an outer product mean of sequence embeddings to the pair embeddings. Conversely, the pair stack affects the MSA stack by biasing its column-wise attention.
We provide a complete description of evoPF in \Crefrange{alg:evopf}{alg:PairTrans} in \Cref{sec:model_arch}.

\subsection{Defining a proper probability distribution over phylogenies with BayesNJ}\label{sec:proper}

After 12 evoPF blocks, we average the MSA embeddings across positions, yielding a single vector embedding per sequence. These embeddings constitute our encoding $\psi(x)$ of the sequences $x$, and our next goal is to define a family of distributions over phylogenies, conditional on $x$, and whose parameters are functions of $\psi(x)$. To this end, we define $q_{\psi(x)}\left(\theta=(\tau,\ell)|x\right)$  factorized over the succession of merges in $\tau$, \emph{i.e.} $q_{\psi(x)}\left(\theta=(\tau,\ell)|x\right)=\Pi_{k=1}^{2N-3}q_m(m^{(k)}|m^{(<k)})q_\ell\left(\ell^{(k)}|m^{(k)},m^{(<k)}\right)$, where $m^{(<k)}$ denotes the set of merges with indices smaller than $k$, and where the probability of each merge is further decomposed into a topological factor $q_m$ representing the probability of merging these two particular clades and a branch-length factor $q_\ell$ factor representing the probability of observing these evolutionary distances between the merged clades and their parents. Both $q_m$ and $q_\ell$ are parameterized by $\psi(x)$ as detailed in~\ref{sec:bayesnj}. A caveat of this factorization across $2N-3$ merges
is that most phylogenies $\theta$ can be obtained by several distinct successions of merges from the leaves. For example, a balanced binary tree with four leaves $a, b, c, d$ merging $(a,b)$ and $(c,d)$ can be obtained by merging any of the two groups first, and these two orders have no reason to lead to the same $\Pi_{k=1}^{2N-3}q_m(m^{(k)}|m^{(<k)})q_\ell\left(\ell^{(k)}|m^{(k)},m^{(<k)}\right)$ in general. Properly defining a probability distribution over phylogenies therefore requires to sum over all possible orders of merge, which is not feasible even for moderately large $N$.

Alternatively, we must ensure that for a given phylogeny our distribution assigns a non-zero probability to a single merge order and, to be able to evaluate the probability of any phylogeny, that this order can be recovered efficiently from the phylogeny. We achieve this by designing a canonical merge order such that (i) we can guarantee that our sampling procedure always generates merges in this order and (ii) our evaluation procedure always processes merges in this order. Point (i) requires that the merge order does not depend on the stochastic parts of the sampling procedure, point (ii) requires that we can recover the canonical order efficiently for any given phylogeny. We achieve these two points by ensuring that at every step $k$, we select the merge corresponding to the two closest nodes currently available in $\mathcal{S}_{(k)}$---\emph{i.e.}, whose children have already been merged. Conversely when sampling a phylogeny $(\tau, \ell)$ from our $q_{\psi(x)}(\tau, \ell|x)$, we ensure that the distance between merged nodes $(i,j)$ is larger than the distance between any previous merges done while $(i,j)$ was a possible merge---\emph{i.e.}, after their children were merged. A similar strategy was used in the likelihood-based sequential Monte Carlo sampler~\cite{bouchard2012phylogenetic}.

\subsection{BayesNJ parameterization of topological and branch-length components}
\label{sec:bayesnj}

We parameterize the topological component of $q_\psi$
by a softmin across pairwise scores computed from a symmetric bilinear function over embeddings of pairs of mergeable nodes at step $k$:
\begin{equation}
	\label{eq:softmin}
	\begin{split}
		 & q_m\left(m^{(k)}|m^{(<k)}\right) = \frac{e^{-\text{score}_{m^{(k)}}}}{\sum_{m'\in\mathcal{C}^{(k)}} e^{-\text{score}_{m'}}},  \text{ and }\forall m = (u,v)\in\mathcal{C}^{(k)}, \text{score}_{m}=\ivec{v}{u}^\top A \ivec{v}{v},
	\end{split}
\end{equation}

where the symmetric matrix $A$ is a learnable parameter and $\ivec{v}{u}$ denotes the embedding for node $u$ in the encoding $\psi(x)$. At $k=1$ these embeddings are those provided by evoPF, and at each successive merge we remove two current leaves and create an embedding for their ancestor which becomes a new leaf (Figure~\ref{fig:PF2}b, Algorithm~\ref{alg:BayesNJ}). The topological component $q_m(m^{(k)}|m^{(<k)})$ is therefore a probabilistic version of the $\argmin$ over pairwise distances that is used in classical hierarchical reconstruction algorithms such as NJ. Furthermore, $q_m$ relies on learnable scores rather than estimates of the evolutionary distances.

In order to ensure that each sampled merge $m^{(k)}$ leads to a longer cherry (sum of branch lengths) than those in all merges previously selected while $m^{(k)}$ was a possible choice---as decided in~\ref{sec:proper} to ensure that $p_{\psi(x)}$ is a proper probability distribution over phylogenies---we re-parameterize the two branch lengths $\left(\ell^{(k)}_i, \ell^{(k)}_j\right) \in \ell^{(k)}$ as their sum $s^{(k)} = \ell^{(k)}_i + \ell^{(k)}_j$ and ratio $r^{(k)} = \ell^{(k)}_i/s^{(k)}$. We update a matrix of constraints as we merge pairs indicating the minimal length that the sum of branch lengths at any new merge must attain for the sampled tree to be consistent with our canonical order. We then model the probability $q_s\left(s^{(k)}|m^{(k)},m^{(<k)}\right)$ as a Gamma distribution shifted by the constraint, and the probability $q_r\left(r^{(k)}|m^{(k)},m^{(<k)}\right)$ of the branch length ratio as a Beta distribution, \emph{i.e.},
\begin{equation}
	\label{eq:bl}
	\begin{split}
		 & q_s\left(s^{(k)}|m^{(\leq k)}\right) = c_{m^{(k)}} + \text{Gamma}\left(\alpha^{(k)}_\text{G}, \lambda^{(k)}_\text{G}\right), \\
		 & q_r\left(r^{(k)}|m^{(\leq k)}\right) = \text{Beta}\left(\alpha^{(k)}_\text{B}, \beta^{(k)}_\text{B}\right),
	\end{split}
\end{equation}

where $c_{m^{(k)}}$ is the current constraint on the sum of branch lengths for merge $m^{(k)}$ and
whose parameters $\left(\alpha^{(k)}_\text{G}, \lambda^{(k)}_\text{G}, \alpha^{(k)}_\text{B}, \beta^{(k)}_\text{B}\right)$ are produced by bilinear functions of the embedding of the two merged nodes $(u,v) = m^{(k)}$---similar to the score used in~\eqref{eq:softmin} for $q_m$. We use a symmetric form for $\left(\alpha^{(k)}_\text{G}, \lambda^{(k)}_\text{G}\right)$ and an asymmetric one for $\left(\alpha^{(k)}_\text{B}, \beta^{(k)}_\text{B}\right)$. The results of the bilinear forms are passed to a softplus function $\text{softplus}(x)=\log\left(1+e^x\right)$ to ensure their positivity, and we further add 1 for all parameters but $\lambda^{(k)}_\text{G}$, as they must be larger than 1. We finally obtain the joint probability $q_\ell\left(\ell^{(k)}|m^{(\leq k)}\right)$ of the two branch lengths as $q_s\left(s^{(k)}|m^{(\leq k)}\right)q_r\left(r^{(k)}|m^{(\leq k)}\right)/s^{(k)}$, where the $1/s^{(k)}$ factor arises from the determinant of the Jacobian of the change of variables from $\left(s^{(k)}, r^{(k)}\right)$ to $\ell^{(k)}$ (see Appendix~\ref{app:jacobian}).

Algorithm~\ref{alg:BayesNJ} summarizes our procedure to evaluate the posterior $q_{\psi(x)}\left(\theta=(\tau,\ell)|x\right)$ of a phylogeny $\tau$ given set of sequences $x$. We use this procedure during the training phase to compute the loss function of Phyloformer~2, and at inference time when we want to evaluate the posterior probability of a phylogeny using a trained network. Of note, when evaluating $q_{\psi(x)}\left(\theta=(\tau,\ell)|x\right)$ the choice of one merge over several possibilities at each step is determined by the data, and does not depend on the network parameters. We therefore don't need to differentiate through discrete operations during training even though the loss evaluation depends on a succession of discrete choices. By contrast, sampling from $q_{\psi(x)}\left(\theta=(\tau,\ell)|x\right)$ does require a sequence of discrete decisions that depend on the network parameters but we never need to differentiate through this process.

Once the network is trained, sampling from the posterior given a set of sequences $x$ (Algorithm~\ref{alg:BayesNJ_sampling}) is very similar to the posterior evaluation procedure described above. The tree is iteratively built from evoPF embeddings of $x$ by successively sampling merges from the softmin-parametrized conditional-merge probabilities~\eqref{eq:softmin}. Similarly, we obtain branch-lengths at a step $k$ by sampling their sum from the shifted Gamma distribution and their ratio from the Beta distribution, whose parameters are obtained from evoPF embeddings~\eqref{eq:bl}.
We can also use the trained network to build a greedy approximation of the maximum \emph{a posteriori} (MAP) estimator (thereafter referred to as greedy MAP), by using the mode of the estimated distributions, \emph{i.e.}, replacing the softmin in~\eqref{eq:softmin} with $\hat{m}_\text{greedyMAP}^{(k)}=\argmin_{m\in\mathcal{C}^{(k)}}\text{score}_m$ to sample the most probable merge at each step $k$, and using the Gamma and Beta modes for branch lengths, \emph{i.e.} $\hat{s}_\text{greedyMAP}^{(k)} = \left(\alpha^{(k)}_\text{G} - 1\right)/\lambda^{(k)}_\text{G}$ and $\hat{r}_\text{greedyMAP}^{(k)}=\left(\alpha^{(k)}_\text{B}-1\right)/\left(\alpha^{(k)}_\text{B}+\beta^{(k)}_\text{B}-2\right)$.

Of note, the resulting $q_\psi$ has limited expressivity for two reasons. First, the relative probabilities for merging each of the currently available pairs are computed based on the initial embeddings (\emph{i.e.}, the embeddings of those nodes that have not been chosen thus far are not updated as the recursion proceeds). In the true posterior, on the other hand, the relative posterior probability of being the next smallest cherry for a given pair of nodes in principle depends on previous merges. An update of the vector of embeddings at each step of the recursion could be implemented in a future version, but would be considerably more computationally intensive.
Second, we model branch lengths posteriors with specific parametric distributions (Gamma and Beta) whereas the true posterior has no reason to match these analytical forms in general.

\section{Experiments}

\subsection{Faster and more accurate point estimates of phylogenies under a tractable model}

We trained Phyloformer~2 (PF2) over a large dataset simulated under similar priors to \citet{nesterenko2025phyloformer}. This dataset contains $\approx 1.3\cdot10^{6}$ 50-taxa trees simulated under a rescaled birth-death process---effectively corresponding to the prior $p(\theta)$---and, for each tree $\theta$, an MSA $x$ simulated under the LG+G8 probabilistic model of evolution $p(x|\theta)$ (LG substitution matrix of~\citet{le_improved_2008} combined with gamma evolution rates~\cite{yang_maximum_1994}, also see \Cref{sec:data}).

The likelihood under this model is tractable, making it a favorable setting for maximum-likelihood methods---FastTree~\citep{price2010FastTree} and IQTREE~\citep{minh2020IQ-TREE} in this experiment. We also include FastME~\citep{lefort2015fastme}, a much faster but less accurate method that only requires to compute likelihoods of branches between pairs of sequences.

The PF2 model was then fine-tuned on tree/MSA pairs with sizes ranging from 10 to 170 taxa, simulated under the same priors as the main training set (see \Cref{sec:data,sec:fine-tuning-batch-size}).
This fine-tuning step is necessary to avoid some overfitting to the number of taxa (see also \Cref{fig:PF2_no_FT}). All MSAs have the same sequence length since overfitting to that metric does not seem to be an issue (see \Cref{fig:MSA_length_overfiting}).

\begin{figure*}[ht]
	\centering
    \ifx\preprint\undefined
    	\includegraphics[width=.8\linewidth]{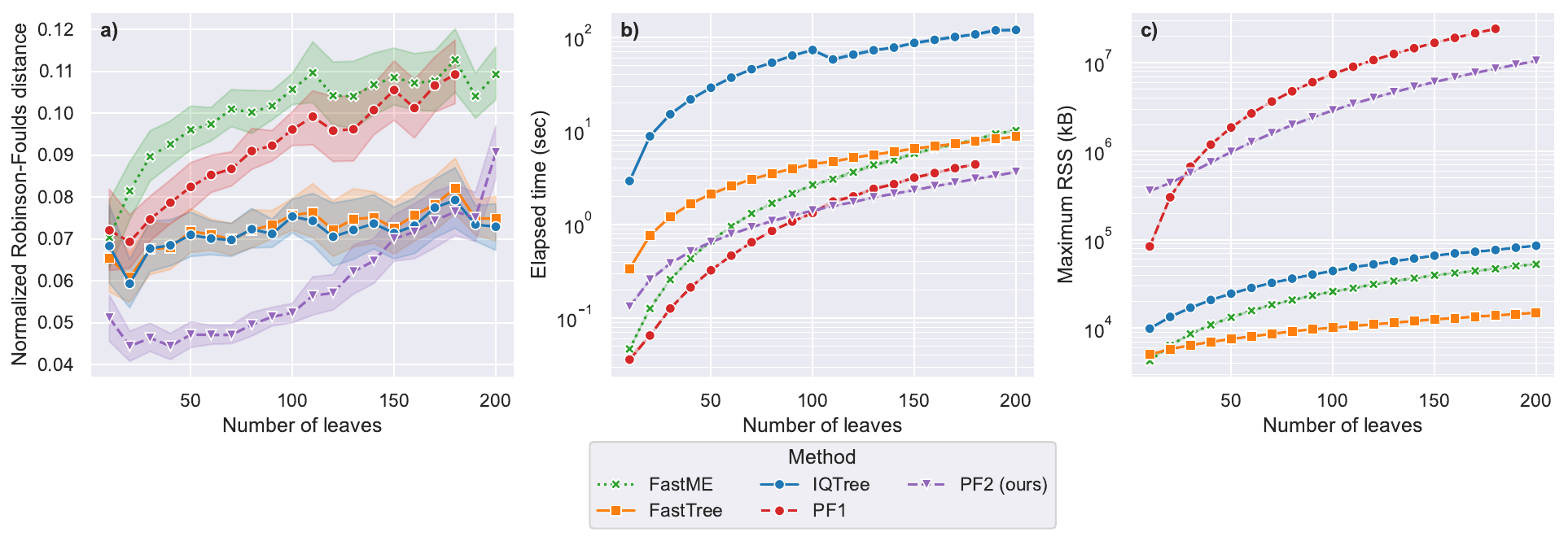}
    \else 
    	\includegraphics[width=.9\linewidth]{figures/PF2_results/ICLR_combined.pdf}
    \fi
	\caption{\textbf{(a)} Topological performance for Phyloformer~2, measured by the normalized Robinson-Foulds distance. The alignments for which trees were inferred were taken from the original Phyloformer paper \citep{nesterenko2025phyloformer} and were simulated under the LG+GC sequence model.
		\textbf{(b)} Runtime and \textbf{(c)} Memory usage for Phyloformer~2.  The same GPU model as the original Phyloformer study was used to run Phyloformer~2 inference.
		Results for other methods are reported from the original Phyloformer paper \citep{nesterenko2025phyloformer}.}
	\label{fig:PF2-LGGC}
\end{figure*}

We then inferred greedy MAP trees (see \Cref{sec:bayesnj}) on a test set of sequence alignments simulated under the same priors as the training set, with 10 to 200 taxa, obtained from~\citet{nesterenko2025phyloformer}.
\Cref{fig:PF2-LGGC}a compares all tested methods using the normalized Robinson-Foulds distance between simulated and inferred tree---a classical metric to compare topologies counting the proportion of branches that are present in only one of two phylogenies~\citep{robinson1981comparison}. PF2 is a marked improvement over PF, with better topological accuracy across the whole range of tree/MSA size. For trees with 10 to 175 leaves, it also reconstructs trees with better accuracies than IQTree and FastTree, both state-of-the-art maximum-likelihood tree reconstruction methods. The edge of PF2 against maximum-likelihood methods working under the correct model likely arises from its estimating the posterior distribution using the correct prior---the same tree distribution is used to generate training and test samples---which reduces its variance without creating a bias.

In addition to being more topologically accurate, PF2 is much faster than maximum-likelihood estimators: by one order of magnitude compared to FastTree, two compared to IQTREE. For trees with more than 100 leaves PF2 is faster than PF. Although PF2 is still memory intensive compared to maximum-likelihood approaches, it scales better than PF allowing PF2 to infer larger trees, faster, despite having 1000 times more parameters than PF (see \Cref{fig:PF2-LGGC}c).

In order to disentangle the effects of the evoPF module and the BayesNJ loss, we trained a $\text{PF2}_{\text{MAE}}$ model using evoPF but replacing the BayesNJ module by a mean absolute error (MAE) loss on pairwise distances, as was done with PF. After a similar number of training steps as PF, we inferred distance matrices from the PF test MSAs, and trees from these matrices using FastME \citep{lefort2015fastme}. $\text{PF2}_{\text{MAE}}$ yields trees with only a slightly better topological accuracy as measured by the RF distance, especially for larger trees (See \Cref{fig:ablation}). This seems to indicate that although the evoPF embedding scheme helps PF2 better predict topologies, most of the topological accuracy gain shown in \Cref{fig:PF2-LGGC}a is due to the BayesNJ loss. Both PF and $\text{PF2}_{\text{MAE}}$ yield similar Kuhner-Felsenstein~\citep[KF, ][]{kuhner1994simulation} distances---a metric similar to RF but weighting each inconsistent branch by the square of its length.

Finally, similar experiments on empirical data are not possible in phylogenetics where the real tree is never observed but we follow~\citet{nesterenko2025phyloformer} and plot for several datasets the distribution of RF distances between the inferred tree for MSAs of individual gene families (obtained with Prank \citep{loytynojaPhylogenyawareAlignmentPRANK2014}) and an estimate of the species tree obtained from a concatenate of gene alignments (\Cref{fig:cyano_rokas}). Since all methods perform inference under the LG model, we expect them to behave similarly---\emph{i.e.}, either all well for datasets where inference under LG leads to a good estimate, or all poorly otherwise. This is what we observe on the cyanobacteria dataset (\Cref{fig:cyano_rokas}a) and three out of the five~\citet{zhouEvaluatingFastMaximum2018} datasets. On the remaing two datasets (NagyA1 and ShenA9, \Cref{fig:cyano_rokas}b), the distribution of errors incurred by PF2 was slightly higher than for other methods. Complementary experiments in~\Cref{app:empirical} clarified that this was caused by the poor quality of some of these empirical alignments which introduced many erroneous gaps. Since PF2 exploits gaps and was trained under a setting where they were consistent with the phylogeny, it is more affected by the issue that
all other methods in the benchmark, which treat gaps as missing data (except PF1). For the same reason, our complementary experiment shows that restoring the phylogenetic signal in the gaps makes PF2 outperform all other methods. 

Besides pointing the sensitivity of PF2 to the alignment quality, these results motivate new developments within PF2 to achieve phylogenetic inference directly from unaligned sequences, which will remove this sensitivity while retaining its unique ability to exploit the insertion-deletion signal.



\subsection{Increased advantage under intractable probabilistic models of evolution}

The main motivation for NPE is to allow well-specified inference under models whose likelihood is intractable. We assess the performance of PF2 in this setting on the same benchmark used in~\citet{nesterenko2025phyloformer}, that includes the Cherry model~\citep{prillo2023cherryml}, allowing for dependencies between evolution at distinct positions in the sequences, and the SelReg model~\citep{duchemin2022evaluation} allowing for heterogeneous amino acid distributions between positions. Of note, a mistake in the generation of the Cherry dataset in~\cite{nesterenko2025phyloformer} inflates the amount of dependency to an unrealistic level (see Appendix~\ref{app:cherry}) but we keep it as it still provides a proof of concept and to avoid the computational burden of re-training PF on a new dataset. We further add a new dataset with sequences evolved under a Potts model~\cite{figliuzzi_how_2018} capturing co-evolutions in the response regulator receiver domain (PF00072) protein family (see Appendix~\ref{sec:data}). 

Under all three models, a fine-tuned PF2 model significantly outperforms the equivalent PF model in RF distance and is comparable to PF in terms of KF score (see
\Cref{fig:pastek-cherry}). For all models and metrics, PF2 (and PF where applicable) outperform all other reference methods. The topological error incurred by PF2 under is typically two or three times lower than the one incurred by likelihood-based methods.

\subsection{Estimating the phylogenetic posterior distribution}


A major advance of PF2 compared to existing likelihood-free phylogenetic inference methods including PF is its ability to represent entire posterior distributions---as opposed to point estimates---over full phylogenies---as opposed to quartets. We assess the quality of this estimation by comparing against samples obtained from a long MCMC run of RevBayes~\citep{hohna2016RevBayes}, a standard tool for Bayesian phylogenetic inference.
We ran 10 parallel MCMC runs on a single 50 sequence alignment for $50,000$ iterations and $5,000$ burn-in iterations. We used a uniform prior on tree topologies and an Exponential distribution with $\lambda=10$ for branch lengths, and LG+G8 as the sequence evolution model. For a fairer comparison, we fine-tuned PF2 on tree/MSA pairs simulated under these priors before sampling from the posterior.

A common way to compare distribution of topologies is through the branches present in sampled trees. Every branch in a phylogeny defines a bipartition of the leaves, making them comparable across all possible trees and providing a softer metric than, \emph{e.g.}, the frequencies of full topologies. \Cref{fig:calibration}a shows that RevBayes produces a hard posterior distribution where most branches appear in either all or none of the sampled topologies. 
PF2 is conservative in its estimate of uncertainty (less certain than it could afford to be), which is expected, given that NPE is trying to minimize the KL divergence from the true distribution to the proposal, which induces a mode-covering behavior, i.e. covers all regions of non-negligible probability mass under the true posterior, at the expense of covering other regions with close to zero posterior probability mass.
PF2 provides a softer posterior but a large agreement with RevBayes, as branches sampled in all RevBayes trees have a frequency larger than 0.6 in PF2, and those not sampled have a frequency mostly lower than 0.3. This is also consistent with our observation that the greedy MAP version of PF2 provides good point estimates, since sequentially sampling the highest probability merges is very likely to select the right branches.

We further investigated the calibration of PF2's posterior probabilities on 196 simulations of trees and alignments with 50 sequences.
We computed branch posterior probabilities from samples of $5,000$ trees sampled from PF2 and compared these probabilities to the frequency with which they correspond to branches in the true simulated tree.
Figure~\ref{fig:calibration}b shows that PF2 is generally well-calibrated, with a tendency to be conservative by generally underestimating branch support.
Overall, PF2 appears to provide  estimates of posterior tree distributions that should be useful in practice. Furthermore it does so several orders of magnitude faster than MCMC methods like RevBayes by leveraging GPU parallelism (see \Cref{sec:break-even}).

\section{Conclusion}

We introduced Phyloformer~2, a phylogenetic neural posterior estimator combining two novel components: evoPF, an expressive and efficient encoding for aligned sequences, and BayesNJ, a factorization of the tree probability over successive merges. Phyloformer~2 can provide MAP estimates that outperforms existing likelihood-based and -free phylogenetic reconstruction methods in topological accuracy while running at least ten times faster. The magnitude of this improvement under models with intractable likelihoods suggests that simplifying assumptions such as homogeneity or site independence that are typically necessary to run likelihood-based inference methods can have a major effect on the quality of phylogenetic reconstruction. The possibility introduced by PF2 of doing inference under more realistic models could therefore be critical in this field.
Furthermore, our experiments show that PF2 approximates posterior distributions in an amortized fashion, with a good calibration, so that branch posterior probabilities provide a useful estimate of their accuracy.


A fundamental challenge in NPE is the potential for distribution shift between the simulated training set and empirical observations. Likelihood-based methods are immune to this issue since they evaluate each observation de novo, without relying on a learning phase. By contrast, NPE may filter out empirical signals that were absent during training but are exploited by maximum likelihood methods. Our results demonstrate that such discrepancies can be diagnosed by monitoring the reconstruction error of empirical embeddings, providing a safeguard against overconfident inference, a necessary feature for the appropriate and safe application of PF2 in real-world settings.


One major limitation of Phyloformer~2 is its scalability, preventing its usage on more than 200 sequences on 16Gb of VRAM. Even though the new inference possibilities under realistic models offered by PF2 are already very valuable for middle-size phylogenies, future work should explore more efficient encoders~\citep{wohlwend2025minifold,wang2025simplefold} or existing heuristics to build larger trees~\citep{warnow2018supertree,jiang2024scaling}.



In the future, we expect Phyloformer~2 to reveal most of its potential by allowing inference under more realistic and complex scenarios~\citep{latrille2021inferring}. Other promising directions are to handle unaligned sequences, and inference under broader probabilistic models of evolution embedding phylogenies such as population dynamics and co-evolution.

\ifx\preprint\undefined
\else
\subsubsection*{Acknowledgments}
This work was funded by the Agence Nationale de la Recherche (ANR-20-CE45-0017). It was granted access to the HPC/AI resources of IDRIS under the allocation AD010714229R2 made by GENCI. The taxon silhouettes in Fig. 1 are modified from public domain images in the PhyloPic database. The authors also thank Vincent Garot and Evan Gorstein.
\fi

\bibliography{bayesnj}

@book{felsenstein2003,
  author = {Felsenstein, J.},
  publisher = {Sinauer Associates},
  title = {Inferring phylogenies},
  year = 2003
}

@article{figliuzzi_how_2018,
	title = {How {Pairwise} {Coevolutionary} {Models} {Capture} the {Collective} {Residue} {Variability} in {Proteins}?},
	volume = {35},
	issn = {0737-4038},
	url = {https://doi.org/10.1093/molbev/msy007},
	doi = {10.1093/molbev/msy007},
	number = {4},
	urldate = {2026-01-29},
	journal = {Molecular Biology and Evolution},
	author = {Figliuzzi, Matteo and Barrat-Charlaix, Pierre and Weigt, Martin},
	month = apr,
	year = {2018},
	pages = {1018--1027},
}

@book{Gelman:2004tc,
author = {Gelman, Andrew and Carlin, John B and Stern, Hal S and Rubin, Donald B},
title = {{Bayesian Data Analysis}},
publisher = {Chapman {\&} Hall/CRC},
year = {2004}
}

@article{le_improved_2008,
	title = {An {Improved} {General} {Amino} {Acid} {Replacement} {Matrix}},
	volume = {25},
	issn = {0737-4038},
	url = {https://doi.org/10.1093/molbev/msn067},
	doi = {10.1093/molbev/msn067},
	number = {7},
	urldate = {2025-02-17},
	journal = {Molecular Biology and Evolution},
	author = {Le, Si Quang and Gascuel, Olivier},
	month = jul,
	year = {2008},
	pages = {1307--1320},
}

@article{hadfield_nextstrain_2018,
	title = {Nextstrain: real-time tracking of pathogen evolution},
	volume = {34},
	issn = {1367-4803},
	shorttitle = {Nextstrain},
	url = {https://doi.org/10.1093/bioinformatics/bty407},
	doi = {10.1093/bioinformatics/bty407},
	number = {23},
	urldate = {2021-03-16},
	journal = {Bioinformatics},
	author = {Hadfield, James and Megill, Colin and Bell, Sidney M and Huddleston, John and Potter, Barney and Callender, Charlton and Sagulenko, Pavel and Bedford, Trevor and Neher, Richard A},
	month = dec,
	year = {2018},
	pages = {4121--4123},
}

@article{10.1111/j.1574-6968.2007.00757.x,
    author = {Aminov, Rustam I. and Mackie, Roderick I.},
    title = {Evolution and ecology of antibiotic resistance genes},
    journal = {FEMS Microbiology Letters},
    volume = {271},
    number = {2},
    pages = {147-161},
    year = {2007},
    month = {06},
    issn = {0378-1097},
    doi = {10.1111/j.1574-6968.2007.00757.x},
    url = {https://doi.org/10.1111/j.1574-6968.2007.00757.x},
    eprint = {https://academic.oup.com/femsle/article-pdf/271/2/147/19604144/271-2-147.pdf},
}

@article{alvarez-carretero_species-level_2022,
	title = {A species-level timeline of mammal evolution integrating phylogenomic data},
	volume = {602},
	copyright = {2021 The Author(s), under exclusive licence to Springer Nature Limited},
	issn = {1476-4687},
	url = {https://www.nature.com/articles/s41586-021-04341-1},
	doi = {10.1038/s41586-021-04341-1},
	language = {en},
	number = {7896},
	urldate = {2022-02-13},
	journal = {Nature},
	author = {Álvarez-Carretero, Sandra and Tamuri, Asif U. and Battini, Matteo and Nascimento, Fabrícia F. and Carlisle, Emily and Asher, Robert J. and Yang, Ziheng and Donoghue, Philip C. J. and dos Reis, Mario},
	month = feb,
	year = {2022},
	note = {Number: 7896
Publisher: Nature Publishing Group},
	pages = {263--267},
}

@article{kapli_phylogenetic_2020,
	title = {Phylogenetic tree building in the genomic age},
	volume = {21},
	copyright = {2020 Springer Nature Limited},
	issn = {1471-0064},
	url = {http://www.nature.com/articles/s41576-020-0233-0},
	doi = {10.1038/s41576-020-0233-0},
	language = {en},
	number = {7},
	urldate = {2025-05-15},
	journal = {Nature Reviews Genetics},
	author = {Kapli, Paschalia and Yang, Ziheng and Telford, Maximilian J.},
	month = jul,
	year = {2020},
	note = {Publisher: Nature Publishing Group},
	pages = {428--444},
}

@article{minh2020IQ-TREE,
    author = {Minh, Bui Quang and Schmidt, Heiko A and Chernomor, Olga and Schrempf, Dominik and Woodhams, Michael D and von Haeseler, Arndt and Lanfear, Robert},
    title = {IQ-TREE 2: New Models and Efficient Methods for Phylogenetic Inference in the Genomic Era},
    journal = {Molecular Biology and Evolution},
    volume = {37},
    number = {5},
    pages = {1530-1534},
    year = {2020},
    month = {02},
    issn = {0737-4038},
    doi = {10.1093/molbev/msaa015},
    url = {https://doi.org/10.1093/molbev/msaa015},
    eprint = {https://academic.oup.com/mbe/article-pdf/37/5/1530/33386032/msaa015.pdf},
}

@article{price2010FastTree,
    doi = {10.1371/journal.pone.0009490},
    author = {Price, Morgan N. AND Dehal, Paramvir S. AND Arkin, Adam P.},
    journal = {PLOS ONE},
    publisher = {Public Library of Science},
    title = {FastTree 2 – Approximately Maximum-Likelihood Trees for Large Alignments},
    year = {2010},
    month = {03},
    volume = {5},
    url = {https://doi.org/10.1371/journal.pone.0009490},
    pages = {1-10},
    number = {3},

}

@article{huelsenbeck2001MRBAYES,
    author = {Huelsenbeck, John P. and Ronquist, Fredrik},
    title = {MRBAYES: Bayesian inference of phylogenetic trees },
    journal = {Bioinformatics},
    volume = {17},
    number = {8},
    pages = {754-755},
    year = {2001},
    month = {08},
    issn = {1367-4803},
    doi = {10.1093/bioinformatics/17.8.754},
    url = {https://doi.org/10.1093/bioinformatics/17.8.754},
    eprint = {https://academic.oup.com/bioinformatics/article-pdf/17/8/754/48837190/bioinformatics\_17\_8\_754.pdf},
}

@article{bouchard2012phylogenetic,
    author = {Bouchard-Côté, Alexandre and Sankararaman, Sriram and Jordan, Michael I.},
    title = {Phylogenetic Inference via Sequential Monte Carlo},
    journal = {Systematic Biology},
    volume = {61},
    number = {4},
    pages = {579-593},
    year = {2012},
    month = {01},
    issn = {1063-5157},
    doi = {10.1093/sysbio/syr131},
    url = {https://doi.org/10.1093/sysbio/syr131},
    eprint = {https://academic.oup.com/sysbio/article-pdf/61/4/579/24563825/syr131.pdf},
}

@inproceedings{zhang2018variational,
title={Variational Bayesian Phylogenetic Inference},
author={Cheng Zhang and Frederick A. Matsen IV},
booktitle={International Conference on Learning Representations},
year={2019},
url={https://openreview.net/forum?id=SJVmjjR9FX},
}

@inproceedings{zhou2024phylogfn,
title={Phylo{GFN}: Phylogenetic inference with generative flow networks},
author={Ming Yang Zhou and Zichao Yan and Elliot Layne and Nikolay Malkin and Dinghuai Zhang and Moksh Jain and Mathieu Blanchette and Yoshua Bengio},
booktitle={The Twelfth International Conference on Learning Representations},
year={2024},
url={https://openreview.net/forum?id=hB7SlfEmze}
}

@inproceedings{duan2024phylogen,
title={PhyloGen: Language Model-Enhanced Phylogenetic Inference via Graph Structure Generation},
author={ChenRui Duan and Zelin Zang and Siyuan Li and Yongjie Xu and Stan Z. Li},
booktitle={The Thirty-eighth Annual Conference on Neural Information Processing Systems},
year={2024},
url={https://openreview.net/forum?id=GxvDsFArxY}
}

@article{hohna2016RevBayes,
    author = {Höhna, Sebastian and Landis, Michael J. and Heath, Tracy A. and Boussau, Bastien and Lartillot, Nicolas and Moore, Brian R. and Huelsenbeck, John P. and Ronquist, Fredrik},
    title = {RevBayes: Bayesian Phylogenetic Inference Using Graphical Models and an Interactive Model-Specification Language},
    journal = {Systematic Biology},
    volume = {65},
    number = {4},
    pages = {726-736},
    year = {2016},
    month = {05},
    issn = {1063-5157},
    doi = {10.1093/sysbio/syw021},
    url = {https://doi.org/10.1093/sysbio/syw021},
    eprint = {https://academic.oup.com/sysbio/article-pdf/65/4/726/16636545/syw021.pdf},
}

@article{felsenstein1981evolutionary,
  title={Evolutionary trees from DNA sequences: a maximum likelihood approach},
  author={Felsenstein, Joseph},
  journal={Journal of molecular evolution},
  volume={17},
  number={6},
  pages={368--376},
  year={1981},
  publisher={Springer}
}

@article{trost2024simulations,
  title={Simulations of sequence evolution: how (un) realistic they are and why},
  author={Trost, Johanna and Haag, Julia and H{\"o}hler, Dimitri and Jacob, Laurent and Stamatakis, Alexandros and Boussau, Bastien},
  journal={Molecular Biology and Evolution},
  volume={41},
  number={1},
  pages={msad277},
  year={2024},
  publisher={Oxford University Press US}
}

@article{telford2005consideration,
	title = {Consideration of {RNA} {Secondary} {Structure} {Significantly} {Improves} {Likelihood}-{Based} {Estimates} of {Phylogeny}: {Examples} from the {Bilateria}},
	volume = {22},
	issn = {0737-4038},
	shorttitle = {Consideration of {RNA} {Secondary} {Structure} {Significantly} {Improves} {Likelihood}-{Based} {Estimates} of {Phylogeny}},
	url = {https://doi.org/10.1093/molbev/msi099},
	doi = {10.1093/molbev/msi099},
	number = {4},
	urldate = {2024-04-12},
	journal = {Molecular Biology and Evolution},
	author = {Telford, Maximilian J and Wise, Michael J and Gowri-Shankar, Vivek},
	month = apr,
	year = {2005},
	pages = {1129--1136}
}

@InProceedings{lueckmann2021benchmarking,
  title = 	 { Benchmarking Simulation-Based Inference },
  author =       {Lueckmann, Jan-Matthis and Boelts, Jan and Greenberg, David and Goncalves, Pedro and Macke, Jakob},
  booktitle = 	 {Proceedings of The 24th International Conference on Artificial Intelligence and Statistics},
  pages = 	 {343--351},
  year = 	 {2021},
  editor = 	 {Banerjee, Arindam and Fukumizu, Kenji},
  volume = 	 {130},
  series = 	 {Proceedings of Machine Learning Research},
  month = 	 {13--15 Apr},
  publisher =    {PMLR},
  url = 	 {https://proceedings.mlr.press/v130/lueckmann21a.html}
}

@article{cranmer2020frontier,
author = {Kyle Cranmer  and Johann Brehmer  and Gilles Louppe },
title = {The frontier of simulation-based inference},
journal = {Proceedings of the National Academy of Sciences},
volume = {117},
number = {48},
pages = {30055-30062},
year = {2020},
doi = {10.1073/pnas.1912789117},
URL = {https://www.pnas.org/doi/abs/10.1073/pnas.1912789117},
eprint = {https://www.pnas.org/doi/pdf/10.1073/pnas.1912789117}}

@InProceedings{greenberg2019automatic,
  title = 	 {Automatic Posterior Transformation for Likelihood-Free Inference},
  author =       {Greenberg, David and Nonnenmacher, Marcel and Macke, Jakob},
  booktitle = 	 {Proceedings of the 36th International Conference on Machine Learning},
  pages = 	 {2404--2414},
  year = 	 {2019},
  editor = 	 {Chaudhuri, Kamalika and Salakhutdinov, Ruslan},
  volume = 	 {97},
  series = 	 {Proceedings of Machine Learning Research},
  month = 	 {09--15 Jun},
  publisher =    {PMLR},
  url = 	 {https://proceedings.mlr.press/v97/greenberg19a.html}
}

@article{radev2020bayesflow,
  title = {{BayesFlow}: Learning complex stochastic models with invertible neural networks},
  author = {Radev, Stefan T. and Mertens, Ulf K. and Voss, Andreas and Ardizzone, Lynton and K{\"o}the, Ullrich},
  journal = {IEEE transactions on neural networks and learning systems},
  volume = {33},
  number = {4},
  pages = {1452--1466},
  year = {2020}
}

@article{suvorov2019accurate,
    author = {Suvorov, Anton and Hochuli, Joshua and Schrider, Daniel R},
    title = "{Accurate Inference of Tree Topologies from Multiple Sequence Alignments Using Deep Learning}",
    journal = {Systematic Biology},
    volume = {69},
    number = {2},
    pages = {221-233},
    year = {2019},
    month = {09},
    issn = {1063-5157},
    doi = {10.1093/sysbio/syz060},
    url = {https://doi.org/10.1093/sysbio/syz060},
    eprint = {https://academic.oup.com/sysbio/article-pdf/69/2/221/38654473/sysbio\_69\_2\_221.pdf},
}

@article{zou2020deep,
  title={Deep residual neural networks resolve quartet molecular phylogenies},
  author={Zou, Zhengting and Zhang, Hongjiu and Guan, Yuanfang and Zhang, Jianzhi},
  journal={Molecular biology and evolution},
  volume={37},
  number={5},
  pages={1495--1507},
  year={2020},
  publisher={Oxford University Press}
}

@article{tang2024novel,
    author = {Tang, Xudong and Zepeda-Nuñez, Leonardo and Yang, Shengwen and Zhao, Zelin and Solís-Lemus, Claudia},
    title = "{Novel symmetry-preserving neural network model for phylogenetic inference}",
    journal = {Bioinformatics Advances},
    volume = {4},
    number = {1},
    pages = {vbae022},
    year = {2024},
    month = {02},
    issn = {2635-0041},
    doi = {10.1093/bioadv/vbae022},
    url = {https://doi.org/10.1093/bioadv/vbae022},
    eprint = {https://academic.oup.com/bioinformaticsadvances/article-pdf/4/1/vbae022/57275075/vbae022.pdf},
}

@article{grosshauser2021reevaluating,
  title={Re-evaluating deep neural networks for phylogeny estimation: the issue of taxon sampling},
  author={Grosshauser, M and Zaharias, P and Warnow T.},
  year={2021},
  journal={Recomb2021},
}

@article{nesterenko2025phyloformer,
    author = {Nesterenko, Luca and Blassel, Luc and Veber, Philippe and Boussau, Bastien and Jacob, Laurent},
    title = {Phyloformer: Fast, Accurate, and Versatile Phylogenetic Reconstruction with Deep Neural Networks},
    journal = {Molecular Biology and Evolution},
    volume = {42},
    number = {4},
    pages = {msaf051},
    year = {2025},
    month = {03},
    issn = {1537-1719},
    doi = {10.1093/molbev/msaf051},
    url = {https://doi.org/10.1093/molbev/msaf051},
    eprint = {https://academic.oup.com/mbe/article-pdf/42/4/msaf051/62373726/msaf051.pdf},
}

@article{saitou1987neighbor,
  title={The neighbor-joining method: a new method for reconstructing phylogenetic trees.},
  author={Saitou, Naruya and Nei, Masatoshi},
  journal={Molecular biology and evolution},
  volume={4},
  number={4},
  pages={406--425},
  year={1987}
}

@article{ho2019axial,
  author    = {Jonathan Ho and
               Nal Kalchbrenner and
               Dirk Weissenborn and
               Tim Salimans},
  title     = {Axial Attention in Multidimensional Transformers},
  journal   = {CoRR},
  volume    = {abs/1912.12180},
  year      = {2019},
  url       = {http://arxiv.org/abs/1912.12180},
  eprinttype = {arXiv},
  eprint    = {1912.12180},
  bibsource = {dblp computer science bibliography, https://dblp.org}
}

@InProceedings{rao2021msatransformer,
  title = 	 {MSA Transformer},
  author =       {Rao, Roshan M and Liu, Jason and Verkuil, Robert and Meier, Joshua and Canny, John and Abbeel, Pieter and Sercu, Tom and Rives, Alexander},
  booktitle = 	 {Proceedings of the 38th International Conference on Machine Learning},
  pages = 	 {8844--8856},
  year = 	 {2021},
  editor = 	 {Meila, Marina and Zhang, Tong},
  volume = 	 {139},
  series = 	 {Proceedings of Machine Learning Research},
  month = 	 {18--24 Jul},
  publisher =    {PMLR},
  url = 	 {https://proceedings.mlr.press/v139/rao21a.html}
}

@misc{katharopoulos2020transformers,
  title={Transformers are rnns: Fast autoregressive transformers with linear attention},
  author={Katharopoulos, Angelos and Vyas, Apoorv and Pappas, Nikolaos and Fleuret, Fran{\c{c}}ois},
  booktitle={International Conference on Machine Learning},
  pages={5156--5165},
  year={2020},
  organization={PMLR}
}

@Article{jumper2021highly,
author={Jumper, John
and Evans, Richard
and Pritzel, Alexander
and Green, Tim
and Figurnov, Michael
and Ronneberger, Olaf
and Tunyasuvunakool, Kathryn
and Bates, Russ
and {\v{Z}}{\'i}dek, Augustin
and Potapenko, Anna
and Bridgland, Alex
and Meyer, Clemens
and Kohl, Simon A. A.
and Ballard, Andrew J.
and Cowie, Andrew
and Romera-Paredes, Bernardino
and Nikolov, Stanislav
and Jain, Rishub
and Adler, Jonas
and Back, Trevor
and Petersen, Stig
and Reiman, David
and Clancy, Ellen
and Zielinski, Michal
and Steinegger, Martin
and Pacholska, Michalina
and Berghammer, Tamas
and Bodenstein, Sebastian
and Silver, David
and Vinyals, Oriol
and Senior, Andrew W.
and Kavukcuoglu, Koray
and Kohli, Pushmeet
and Hassabis, Demis},
title={Highly accurate protein structure prediction with AlphaFold},
journal={Nature},
year={2021},
month={Aug},
day={01},
volume={596},
number={7873},
pages={583-589},
issn={1476-4687},
doi={10.1038/s41586-021-03819-2},
url={https://doi.org/10.1038/s41586-021-03819-2}
}

@article{lefort2015fastme,
  title={FastME 2.0: a comprehensive, accurate, and fast distance-based phylogeny inference program},
  author={Lefort, Vincent and Desper, Richard and Gascuel, Olivier},
  journal={Molecular biology and evolution},
  volume={32},
  number={10},
  pages={2798--2800},
  year={2015},
  publisher={Oxford University Press}
}

@misc{wang2025simplefold,
      title={SimpleFold: Folding Proteins is Simpler than You Think}, 
      author={Yuyang Wang and Jiarui Lu and Navdeep Jaitly and Josh Susskind and Miguel Angel Bautista},
      year={2025},
      eprint={2509.18480},
      archivePrefix={arXiv},
      primaryClass={cs.LG},
      url={https://arxiv.org/abs/2509.18480}, 
}

@article{wohlwend2025minifold,
title={MiniFold: Simple, Fast, and Accurate Protein Structure Prediction},
author={Jeremy Wohlwend and Mateo Reveiz and Matt McPartlon and Axel Feldmann and Wengong Jin and Regina Barzilay},
journal={Transactions on Machine Learning Research},
issn={2835-8856},
year={2025},
url={https://openreview.net/forum?id=1p9hQTbjgo},
note={Featured Certification}
}

@article{latrille2021inferring,
    author = {Latrille, Thibault and Lanore, Vincent and Lartillot, Nicolas},
    title = {Inferring Long-Term Effective Population Size with Mutation–Selection Models},
    journal = {Molecular Biology and Evolution},
    volume = {38},
    number = {10},
    pages = {4573-4587},
    year = {2021},
    month = {06},
    issn = {1537-1719},
    doi = {10.1093/molbev/msab160},
    url = {https://doi.org/10.1093/molbev/msab160},
    eprint = {https://academic.oup.com/mbe/article-pdf/38/10/4573/40449557/msab160.pdf},
}

@ARTICLE{jiang2024scaling,
  title    = "Scaling {DEPP} phylogenetic placement to ultra-large reference
              trees: a tree-aware ensemble approach",
  author   = "Jiang, Yueyu and McDonald, Daniel and Perry, Daniela and Knight,
              Rob and Mirarab, Siavash",
  journal  = "Bioinformatics",
  volume   =  40,
  number   =  6,
  month    =  jun,
  year     =  2024,
  address  = "England",
  language = "en"
}

@misc{warnow2018supertree,
      title={Supertree Construction: Opportunities and Challenges}, 
      author={Tandy Warnow},
      year={2018},
      eprint={1805.03530},
      archivePrefix={arXiv},
      primaryClass={q-bio.PE},
      url={https://arxiv.org/abs/1805.03530}, 
}

@article{robinson1981comparison,
title = {Comparison of phylogenetic trees},
journal = {Mathematical Biosciences},
volume = {53},
number = {1},
pages = {131-147},
year = {1981},
issn = {0025-5564},
doi = {https://doi.org/10.1016/0025-5564(81)90043-2},
url = {https://www.sciencedirect.com/science/article/pii/0025556481900432},
author = {D.F. Robinson and L.R. Foulds}
}

@article{duchemin2022evaluation,
    author = {Duchemin, Louis and Lanore, Vincent and Veber, Philippe and Boussau, Bastien},
    title = {Evaluation of Methods to Detect Shifts in Directional Selection at the Genome Scale},
    journal = {Molecular Biology and Evolution},
    volume = {40},
    number = {2},
    pages = {msac247},
    year = {2022},
    month = {12},
    issn = {1537-1719},
    doi = {10.1093/molbev/msac247},
    url = {https://doi.org/10.1093/molbev/msac247},
    eprint = {https://academic.oup.com/mbe/article-pdf/40/2/msac247/49278916/msac247.pdf},
}

@Article{prillo2023cherryml,
author={Prillo, Sebastian
and Deng, Yun
and Boyeau, Pierre
and Li, Xingyu
and Chen, Po-Yen
and Song, Yun S.},
title={CherryML: scalable maximum likelihood estimation of phylogenetic models},
journal={Nature Methods},
year={2023},
month={Aug},
day={01},
volume={20},
number={8},
pages={1232-1236},
issn={1548-7105},
doi={10.1038/s41592-023-01917-9},
url={https://doi.org/10.1038/s41592-023-01917-9}
}

@ARTICLE{kuhner1994simulation,
  title    = "A simulation comparison of phylogeny algorithms under equal and
              unequal evolutionary rates",
  author   = "Kuhner, M K and Felsenstein, J",
  journal  = "Mol Biol Evol",
  volume   =  11,
  number   =  3,
  pages    = "459--468",
  month    =  may,
  year     =  1994,
  address  = "United States",
  language = "en"
}

@inproceedings{koptagel2022vaiphy,
 author = {Koptagel, Hazal and Kviman, Oskar and Melin, Harald and Safinianaini, Negar and Lagergren, Jens},
 booktitle = {Advances in Neural Information Processing Systems},
 editor = {S. Koyejo and S. Mohamed and A. Agarwal and D. Belgrave and K. Cho and A. Oh},
 pages = {14758--14770},
 publisher = {Curran Associates, Inc.},
 title = {VaiPhy: a Variational Inference Based Algorithm for Phylogeny},
 url = {https://proceedings.neurips.cc/paper_files/paper/2022/file/5e956fef0946dc1e39760f94b78045fe-Paper-Conference.pdf},
 volume = {35},
 year = {2022}
}

@article{yang_maximum_1994,
	title = {Maximum likelihood phylogenetic estimation from {DNA} sequences with variable rates over sites: approximate methods},
	volume = {39},
	language = {eng},
	number = {3},
	journal = {Journal of Molecular Evolution},
	author = {Yang, Z.},
	month = sep,
	year = {1994},
	pages = {306--14},
}

@online{pagnaniGenerativeContinuousTime2025a,
  title = {Generative Continuous Time Model Reveals Epistatic Signatures in Protein Evolution},
  author = {Pagnani, Andrea and Barrat-Charlaix, Pierre},
  date = {2025-09-20},
  eprinttype = {bioRxiv},
  eprintclass = {New Results},
  pages = {2025.09.17.676821},
  issn = {2692-8205},
  doi = {10.1101/2025.09.17.676821},
  url = {https://www.biorxiv.org/content/10.1101/2025.09.17.676821v1},
  urldate = {2026-01-11},
  langid = {english},
  pubstate = {prepublished}
}

@dataset{szollosiDataEfficientExploration2013,
  title = {Data from: {{Efficient}} Exploration of the Space of Reconciled Gene Trees},
  shorttitle = {Data From},
  author = {Szöllősi, Gergely J. and Rosikiewicz, Wojciech and Boussau, Bastien and Tannier, Eric and Daubin, Vincent},
  date = {2013-08-07},
  pages = {45688314 bytes},
  publisher = {[object Object]},
  doi = {10.5061/DRYAD.PV6DF},
  url = {https://datadryad.org/stash/dataset/doi:10.5061/dryad.pv6df},
  langid = {english},
  version = {1}
}

@article{zhouEvaluatingFastMaximum2018,
  title = {Evaluating {{Fast Maximum Likelihood-Based Phylogenetic Programs Using Empirical Phylogenomic Data Sets}}},
  author = {Zhou, Xiaofan and Shen, Xing-Xing and Hittinger, Chris Todd and Rokas, Antonis},
  date = {2018-02-01},
  journaltitle = {Molecular Biology and Evolution},
  shortjournal = {Molecular Biology and Evolution},
  volume = {35},
  number = {2},
  pages = {486--503},
  issn = {0737-4038},
  doi = {10.1093/molbev/msx302},
  url = {https://doi.org/10.1093/molbev/msx302},
}

@article{mistryPfamProteinFamilies2021a,
  title = {Pfam: {{The}} Protein Families Database in 2021},
  shorttitle = {Pfam},
  author = {Mistry, Jaina and Chuguransky, Sara and Williams, Lowri and Qureshi, Matloob and Salazar, Gustavo A and Sonnhammer, Erik L L and Tosatto, Silvio C E and Paladin, Lisanna and Raj, Shriya and Richardson, Lorna J and Finn, Robert D and Bateman, Alex},
  date = {2021-01-01},
  journaltitle = {Nucleic acids research},
  shortjournal = {Nucleic Acids Res},
  volume = {49},
  number = {D1},
  eprint = {33125078},
  eprinttype = {pubmed},
  pages = {D412-D419},
  issn = {1362-4962},
  doi = {10.1093/nar/gkaa913},
  url = {https://europepmc.org/articles/PMC7779014},
  langid = {english},
  pmcid = {PMC7779014},
}

@incollection{rossetAdabmDCA20AFlexible2026,
  title = {{{adabmDCA}} 2.0—{{A Flexible}} but {{Easy-to-Use Package}} for {{Direct Coupling Analysis}}},
  booktitle = {Protein {{Evolution}}: {{Methods}} and {{Protocols}}},
  author = {Rosset, Lorenzo and Netti, Roberto and Muntoni, Anna Paola and Weigt, Martin and Zamponi, Francesco},
  editor = {Khan, Shahid M. and Pazos, Florencio},
  date = {2026},
  pages = {83--104},
  publisher = {Springer US},
  location = {New York, NY},
  doi = {10.1007/978-1-0716-4828-5_6},
  url = {https://doi.org/10.1007/978-1-0716-4828-5_6},
  urldate = {2026-01-29},
  isbn = {978-1-0716-4828-5},
  langid = {english}
}

@incollection{loytynojaPhylogenyawareAlignmentPRANK2014,
  title = {Phylogeny-Aware Alignment with {{PRANK}}},
  booktitle = {Multiple {{Sequence Alignment Methods}}},
  author = {Löytynoja, Ari},
  editor = {Russell, David J},
  date = {2014},
  pages = {155--170},
  publisher = {Humana Press},
  location = {Totowa, NJ},
  doi = {10.1007/978-1-62703-646-7_10},
  url = {https://doi.org/10.1007/978-1-62703-646-7_10},
  urldate = {2026-05-05},
  isbn = {978-1-62703-646-7},
  langid = {english}
}

@article{katoh_multiple_2013,
    title = {Multiple {Sequence} {Alignment} {Software} {Version} 7: {Improvements} in performance and usability.},
    volume = {30},
    shorttitle = {Katoh {K}, {Standley} {DM}.. {MAFFT} {Multiple} {Sequence} {Alignment} {Software} {Version} 7},
    url = {https://academic.oup.com/mbe/article/30/4/772/1073398},
    doi = {10.1093/molbev/mst010},
    journal = {Molecular biology and evolution},
    author = {Katoh, Kazutaka and Standley, Daron},
    month = jan,
    year = {2013},
}

@article{wong2008alignment,
author = {Karen M. Wong  and Marc A. Suchard  and John P. Huelsenbeck },
title = {Alignment Uncertainty and Genomic Analysis},
journal = {Science},
volume = {319},
number = {5862},
pages = {473-476},
year = {2008},
doi = {10.1126/science.1151532},
URL = {https://www.science.org/doi/abs/10.1126/science.1151532},
eprint = {https://www.science.org/doi/pdf/10.1126/science.1151532}}

@article{loytynoja2008phylogeny,
author = {Ari Löytynoja  and Nick Goldman },
title = {Phylogeny-Aware Gap Placement Prevents Errors in Sequence Alignment and Evolutionary Analysis},
journal = {Science},
volume = {320},
number = {5883},
pages = {1632-1635},
year = {2008},
doi = {10.1126/science.1158395},
URL = {https://www.science.org/doi/abs/10.1126/science.1158395},
eprint = {https://www.science.org/doi/pdf/10.1126/science.1158395}}
\bibliographystyle{icml2026}

\section*{Code and Data Availability}
All the code used in this study is available at the following anonymized repository: 
\href{https://anonymous.4open.science/r/Phyloformer2-EDCF/README.md}{anonymous.4open.science/r/Phyloformer2-EDCF}

\newpage
\appendix
\onecolumn

\section{Technical Appendices and Supplementary Material}
Technical appendices with additional results, figures, graphs and proofs may be submitted with the paper submission before the full submission deadline (see above), or as a separate PDF in the ZIP file below before the supplementary material deadline. There is no page limit for the technical appendices.

\subsection{Model architecture}
\label{sec:model_arch}

\paragraph{Notation:} $i$, $j$ and $k$ denote sequence indices, $t$ and $l$ denote residue indices. Capitalized functions (e.g. $\lin$) are parametrized and learnable, whereas lowercase functions (e.g. $\sig$) are not.

The evoPF module takes as input a multiple sequence alignments $\{\ivec{x}{it}\}$, where $\ivec{x}{it}$ is the one-hot encoded vector of the $t^{th}$ character of sequence $i$.
The evoPF module produces a set sequence embeddings $\{\ivec{v}{i}\}$ of fixed size $c_s=128$ for each sequence $i$, and a set a sequence pair embeddings $\{\ivec{z}{ij}\}$ of fixed size $c_z=256$ for each pair $(i,j)$ of sequences. 

\begin{algorithm}[H]
	\caption{The evoPF module}
	\label{alg:evopf}
	\begin{algorithmic}
    \FUNCTION{\textsc{EvoPF} ($N_{blocs}=12$, $\{\ivec{x}{it}\}$)}
		\STATE $\{\ivec{v}{it}\}$ = MSAEmbedder$(\{\ivec{x}{it}\})$
		\STATE $\{\ivec{z}{ij}\}$ = PairEmbedder$(\{\ivec{x}{it}\})$
		\\
		\FOR{$l\in[1,\ldots,N_{blocs}]$}
		\STATE $\{\ivec{v}{it}\}$ += MSAColAttentionWithPairBias$(\{\ivec{v}{it}\}, \{\ivec{z}{ij}\})$
		\STATE $\{\ivec{v}{it}\}$ += MSARowAttention$(\{\ivec{v}{it}\})$
		\STATE $\{\ivec{v}{it}\}$ += MSATransition$(\{\ivec{v}{it}\})$
		\\
		\STATE $\{\ivec{z}{ij}\}$ += OuterProductMean$(\{\ivec{v}{it}\})$
		\\
		\STATE $\{\ivec{z}{ij}\}$ += PairAttention$(\{\ivec{z}{ij}\})$
		\STATE $\{\ivec{z}{ij}\}$ += PairTransition$(\{\ivec{z}{ij}\})$
		\ENDFOR
		\\
		\STATE $z_{ij} = \lin(\ivec{z}{ij})$
		\STATE $\ivec{v}{i} = \mean_t(\ivec{v}{it})$
		\STATE \textbf{return} $\{z_{ij}\}, \{\ivec{v}{i}\}$\COMMENT{$\ivec{z}{ij}\in\mathbb{R}^{c_z}$, $\ivec{v}{i}\in\mathbb{R}^{c_s}$}
		\ENDFUNCTION
	\end{algorithmic}
\end{algorithm}

\subsubsection{Embedding modules}

\begin{algorithm}[H]
	\caption{MSA Embedding module}    \label{alg:msaembed}
	\begin{algorithmic}
    \FUNCTION{\textsc{MSAEmbedder} ($\{\ivec{x}{it}\}$, $c_s=128$)}
		\STATE $\ivec{a}{it} = \conv(\ivec{x}{it})$ \COMMENT{$a_{it}\in\mathbb{R}^{c_s}$}
		\STATE $\ivec{v}{it} = \relu(\ivec{a}{it})$
		\STATE $\{\ivec{v}{it}\}$
		\ENDFUNCTION
	\end{algorithmic}
\end{algorithm}

\begin{algorithm}[H]
	\caption{Pair Embedding module}    \label{alg:pairembed}
	\begin{algorithmic}
    \FUNCTION{\textsc{PairEmbedder} ($\{\ivec{x}{it}\}$, $c_z=256$)}
		\STATE $a_{it} = \relu(\conv(\ivec{x}{it})$) \COMMENT{$a_{it}\in\mathbb{R}^{c_z}$}
			\STATE $\ivec{z}{ij} = \mean_t(a_{it} + a_{jt})$ \COMMENT{$1\le j<i\le N$}
			\STATE \textbf{return} $\{\ivec{z}{ij}\}$
		\ENDFUNCTION
	\end{algorithmic}
\end{algorithm}

\subsubsection{The evoPF MSA stack}
\begin{algorithm}[H]
	\caption{MSA Stack - Column-wise pair-biased gated self-attention}\label{alg:ColAtt}
	\begin{algorithmic}
    \FUNCTION{\textsc{MSAColAttentionWithPairBias} ($\{\ivec{v}{it}\}, \{\ivec{z}{ij}\}, N_{head}=4,c=c_s/N_{head}$)}
		\STATE $\ivec{v}{it}\gets\lnorm(\ivec{v}{it})$
		\STATE $\ivec{q}{it}^h,\ivec{t}{it}^h,\ivec{v}{it}^h, = \linn(\ivec{v}{it})$ \COMMENT{$\ivec{q}{it}^h,\ivec{t}{it}^h,\ivec{v}{it}^h\in \mathbb{R}^c,1\le h\le N_{head}$}
		\STATE $b_{ij}^h = \linn(\ivec{z}{ij})$
		\STATE $\ivec{g}{it}^h = \sig(\lin(\ivec{v}{it})$) \COMMENT{$\ivec{g}{it}\in\mathbb{R}^c$}
			\\
			\STATE $a_{ijt}^h = \smx\left(\frac{1}{\sqrt{c}}\ivec{q}{it}^{h\top} \ivec{t}{jt}^h + b_{ij}^h\right)$
			\STATE $\ivec{o}{it}^h = \ivec{g}{it}^h \odot \sum_j a_{ijt}^h\ivec{v}{it}^h$
			\\
			\STATE $\ivec{\tilde{s}}{it} = \lin\left(\ct_h(\ivec{o}{it}^h)\right)$ \COMMENT{$\ivec{\tilde{s}}{it}\in\mathbb{R}^{c_s}$}
			\STATE\textbf{return} $\{\ivec{\tilde{s}}{it}\}$
		\ENDFUNCTION
	\end{algorithmic}
\end{algorithm}

\begin{algorithm}[H]
	\caption{MSA Stack - Row-wise gated self-attention}\label{alg:RowAtt}
	\begin{algorithmic}
    \FUNCTION{\textsc{MSARowAttention} ($\{\ivec{v}{it}\}, N_{head}=4,c=c_s/N_{head}$)}
		\STATE $\ivec{v}{it}\gets\lnorm(\ivec{v}{it})$
		\STATE $\ivec{q}{it}^h,\ivec{t}{it}^h,\ivec{v}{it}^h, = \linn(\ivec{v}{it})$ \COMMENT{$\ivec{q}{it}^h,\ivec{t}{it}^h,\ivec{v}{it}^h\in \mathbb{R}^c,1\le h\le N_{head}$}
		\STATE $\ivec{g}{it}^h = \sig(\lin(\ivec{v}{it})$) \COMMENT{$\ivec{g}{it}\in\mathbb{R}^c$}
			\\
			\STATE $a_{itl}^h = \smx\left(\frac{1}{\sqrt{c}}\ivec{q}{it}^{h\top} \ivec{t}{il}^h\right)$
			\STATE $\ivec{o}{it}^h = \ivec{g}{it}^h \odot \sum_l a_{itl}^h\ivec{v}{il}^h$
			\\
			\STATE $\ivec{\tilde{s}}{it} = \lin\left(\ct_h(\ivec{o}{it}^h)\right)$ \COMMENT{$\ivec{\tilde{s}}{it}\in\mathbb{R}^{c_s}$}
			\STATE\textbf{return} $\{\ivec{\tilde{s}}{it}\}$
		\ENDFUNCTION
	\end{algorithmic}
\end{algorithm}

\begin{algorithm}[H]
	\caption{MSA Stack - Transition}\label{alg:MSAtrans}
	\begin{algorithmic}
    \FUNCTION{\textsc{MSATransition} ($\{\ivec{v}{it}\}, n=4$)}
		\STATE $\ivec{v}{it}\gets\lnorm(\ivec{v}{it})$
		\STATE $\ivec{a}{it} = \lin(\ivec{v}{it})$\COMMENT{$\ivec{a}{it}\in\mathbb{R}^{nc_s}$}
		\STATE $\ivec{v}{it} \gets \lin(\relu(\ivec{a}{it}))$
		\STATE\textbf{return} $\{\ivec{v}{it}\}$
		\ENDFUNCTION
	\end{algorithmic}
\end{algorithm}

\begin{algorithm}[H]
	\caption{Communication - Outer product mean}\label{alg:OPM}
	\begin{algorithmic}
    \FUNCTION{\textsc{OuterProductMean} ($\{\ivec{v}{ik}\}, c=32$)}
		\STATE $\ivec{v}{it}\gets\lnorm(\ivec{v}{it})$
		\STATE $\ivec{a}{it},\ivec{b}{it} = \lin(\ivec{v}{it})$\COMMENT{$\ivec{a}{it},\ivec{b}{it}\in\mathbb{R}^{nc_s}$}
		\STATE $\ivec{o}{ij} \gets \text{flatten}(\mean_t(\ivec{a}{it}\otimes\ivec{b}{it}))$\COMMENT{$\ivec{o}{ij}\in\mathbb{R}^{c^2}$}
		\STATE $\ivec{z}{ij} = \lin(\ivec{o}{ij})$ \COMMENT{$\ivec{z}{ij}\in\mathbb{R}^{c_z}$}
		\STATE\textbf{return} $\{\ivec{z}{ij}\}$
		\ENDFUNCTION
	\end{algorithmic}
\end{algorithm}

\subsubsection{The evoPF pair stack}
\begin{algorithm}[H]
	\caption{Pair Stack - Gated self-attention}\label{alg:PairAtt}
	\begin{algorithmic}
    \FUNCTION{\textsc{PairAttention} ($\{\ivec{z}{ij}\}, N_{head}=4,c=c_z/N_{head}$)}
		\STATE $\ivec{z}{ij}\gets\lnorm(\ivec{z}{ij})$
		\STATE $\ivec{q}{ij}^h,\ivec{k}{ij}^h,\ivec{v}{ij}^h, = \linn(\ivec{z}{ij})$ \COMMENT{$\ivec{q}{ik}^h,\ivec{k}{ij}^h,\ivec{v}{ij}^h\in \mathbb{R}^c,1\le h\le N_{head}$}
		\STATE $\ivec{g}{ij}^h = \sig(\lin(\ivec{z}{ij})$) \COMMENT{$\ivec{g}{ij}\in\mathbb{R}^c$}
			\\
			\STATE $a_{ijk}^h = \smx\left(\frac{1}{\sqrt{c}}\ivec{q}{ij}^{h\top} \ivec{k}{jk}^h\right)$
			\STATE $\ivec{o}{ij}^h = \ivec{g}{ij}^h \odot \sum_k a_{ijk}^h\ivec{v}{ik}^h$
			\\
			\STATE $\ivec{\tilde{z}}{ij} = \lin\left(\ct_h(\ivec{o}{ij}^h)\right)$ \COMMENT{$\ivec{\tilde{z}}{ik}\in\mathbb{R}^{c_z}$}
			\STATE\textbf{return} $\{\ivec{\tilde{z}}{ij}\}$
		\ENDFUNCTION
	\end{algorithmic}
\end{algorithm}

\begin{algorithm}
	\caption{Pair Stack - Transition}\label{alg:PairTrans}
	\begin{algorithmic}
    \FUNCTION{\textsc{PairTransition} ($\{\ivec{z}{ij}\}, n=4$)}
		\STATE $\ivec{z}{ij}\gets\lnorm(\ivec{z}{ij})$
		\STATE $\ivec{a}{ij} = \lin(\ivec{z}{ij})$\COMMENT{$\ivec{a}{ij}\in\mathbb{R}^{nc_z}$}
		\STATE $\ivec{z}{ij} \gets \lin(\relu(\ivec{a}{ij}))$
		\STATE\textbf{return} $\{\ivec{z}{ij}\}$
		\ENDFUNCTION
	\end{algorithmic}
\end{algorithm}

\newpage
\subsubsection{Evaluating $q_{\psi(x)}\left(\theta=(\tau,\ell)|x\right)$ with BayesNJ}

\begin{algorithm}[ht!]
	\caption{The BayesNJ Loss}
	\label{alg:BayesNJ}
	\begin{algorithmic}
    \FUNCTION{\textsc{BayesNJ} (Embeddings $\{\psi(x)\} = \{\ivec{v}{1},\ldots,\ivec{v}{N}\}$, merges $\{m^{(k)}\}$, branch lengths $\{\ell^{(k)}\}$)}
		\STATE $\mathcal{L}_{\tau}\gets0, \mathcal{L}_{\ell}\gets0, \{c^{(1)}_{m}\}\gets\{0\, \forall m \in\mathcal{S}_{(1)}\}$
		\COMMENT{Initialize log probabilities and constraints}
		\FORALL{$k \in [1,\dots,N-2]$}
		\STATE \textcolor{gray}{\# Topological component}
		\STATE $\{\text{score}_{m=(u,v)}\} \gets \left\{\text{SymmetricBilinear}(\ivec{v}{u},\ivec{v}{v})\right\}$\COMMENT{Score all pairs $m=(u,v)\in\mathcal{C}{(k)}$}
		\STATE $\{q_{m}(m)\} \gets \smn(\{\text{score}_{m}\})$
		\STATE $\mathcal{L}_\tau \mathrel{+}= \log(q_{m}(m^{(k)})$
		\\\medskip
		\STATE \textcolor{gray}{\# Branch length component}
		\STATE $\ivec{v}{N+k} \gets \text{SymmetricBilinear}(m^{(k)})$ \COMMENT{Compute parent embedding}
		\STATE $\left(\tilde{\alpha}^{(k)}_\text{G}, \tilde{\lambda}^{(k)}_\text{G}\right) \gets  \text{SymmetricBilinear}(m^{(k)})$
		\STATE $\tilde{\alpha}^{(k)}_\text{G} \gets 1 + \tilde{\alpha}^{(k)}_\text{G}$
		\STATE $\left(\alpha^{(k)}_\text{G}, \lambda^{(k)}_\text{G}\right) \gets  \text{softplus}\left(\tilde{\alpha}^{(k)}_\text{G}, \tilde{\lambda}^{(k)}_\text{G}\right) + \varepsilon$
		\STATE $\left(\tilde{\alpha}^{(k)}_\text{B}, \tilde{\beta}^{(k)}_\text{B}\right) \gets  \text{Bilinear}(m^{(k)})$
		\STATE $\left(\alpha^{(k)}_\text{B}, \beta^{(k)}_\text{B}\right) \gets 1 + \text{softplus}\left(\tilde{\alpha}^{(k)}_\text{B}, \tilde{\beta}^{(k)}_\text{B}\right) + \varepsilon$
		\STATE $\mathcal{L}_\ell \mathrel{+}= \log\text{PDF}_{\text{Gamma}}\left(\ell_i^{(k)} + \ell_j^{(k)} - c^{(k)}_{m^{(k)}}|\alpha^{(k)}_\text{G}, \lambda^{(k)}_\text{G}\right) + \log\text{PDF}_{\text{Beta}}\left(\ell_i^{(k)}/\ell_j^{(k)}|\alpha^{(k)}_\text{B}, \beta^{(k)}_\text{B}\right)$
		\\\medskip
		\STATE $\{c^{(k+1)}_{m}\} \gets \left\{\max\left(c^{(k)}_{m}, \ell_i^{(k)}+\ell_j^{(k)}\right),
			\forall m\in\mathcal{S}_{(k)}\right\}$
		\COMMENT{Update constraints}
    \\\medskip
		\STATE \textcolor{gray}{\# Prepare next iteration}
		\STATE $\mathcal{S}_{(k+1)} = \left\{\mathcal{S}_{(k)}\cup u\right\}\setminus m^{(k)}$ \COMMENT{Update mergeable nodes}
		\ENDFOR
		\STATE \textbf{return} $\mathcal{L}_{\tau}, \mathcal{L}_{\ell}$
		\ENDFUNCTION
	\end{algorithmic}
\end{algorithm}

\newpage

\subsubsection{Sampling from $q_{\psi(x)}\left(\theta=(\tau,\ell)|x\right)$ with BayesNJ}
\label{sec:suppBayesNJ}

\begin{algorithm}[H]
	\caption{Sampling from the posterior\\
		For the greedy MAP approximation, we simply replace the softmin of scores with an argmin to sample merges, and replace samplings from Gamma and Beta distributions with the corresponding mode.}
	\label{alg:BayesNJ_sampling}
	\begin{algorithmic}
    \FUNCTION{\textsc{SamplingBayesNJ} (Embeddings $\{\psi(x)\} = \{\ivec{v}{1},\ldots,\ivec{v}{N}\}$)}
		\STATE $\{c^{(1)}_{ij}\}\gets\{0,\forall(i,j)\in[1,\dots,N-2]^2\}$ \COMMENT{Initialize constraints}
		\FORALL{$k \in [1,\dots,N-2]$}
		\STATE \textcolor{gray}{\# Topological component}
		\STATE $\{\text{score}_{ij}\} \gets \left\{\text{SymmetricBilinear}(\ivec{v}{i},\ivec{v}{j})\right\}$\COMMENT{Score all pairs $(i,j)\in\mathcal{S}_{(k)}^2$}
		\STATE $m^{(k)} \sim \smn(\{\text{score}_{ij}\})$\COMMENT{Sample merge}
		\\\medskip
		\STATE \textcolor{gray}{\# Branch length component}
		\STATE $\ivec{v}{N+k} \gets \text{SymmetricBilinear}(m^{(k)})$ \COMMENT{Compute parent embedding}
		\STATE $\left(\tilde{\alpha}^{(k)}_\text{G}, \tilde{\lambda}^{(k)}_\text{G}\right) \gets  \text{SymmetricBilinear}(m^{(k)})$
		\STATE $\tilde{\alpha}^{(k)}_\text{G} \gets 1 + \tilde{\alpha}^{(k)}_\text{G}$
		\STATE $\left(\alpha^{(k)}_\text{G}, \lambda^{(k)}_\text{G}\right) \gets  \text{softplus}\left(\tilde{\alpha}^{(k)}_\text{G}, \tilde{\lambda}^{(k)}_\text{G}\right) + \varepsilon$
		\STATE $\left(\tilde{\alpha}^{(k)}_\text{B}, \tilde{\beta}^{(k)}_\text{B}\right) \gets  \text{Bilinear}(m^{(k)})$
		\STATE $\left(\alpha^{(k)}_\text{B}, \beta^{(k)}_\text{B}\right) \gets 1 + \text{softplus}\left(\tilde{\alpha}^{(k)}_\text{B}, \tilde{\beta}^{(k)}_\text{B}\right) + \varepsilon$

		\STATE $s^{(k)} \sim c^{(k)}_{m^{(k)}} + \text{Gamma}(\alpha^{(k)}_\text{G}, \lambda^{(k)}_\text{G})$\COMMENT{Sample sum}
		\STATE $r^{(k)}\sim\text{Beta}(\alpha^{(k)}_\text{B}, \beta^{(k)}_\text{B})$\COMMENT{Sample ratio}

		\STATE$\ell^{(k)}_i\gets r^{(k)}\cdot s^{(k)}$
		\STATE$\ell^{(k)}_j\gets s^{(k)} - \ell^{(k)}_i$
    \\\medskip
		\STATE \textcolor{gray}{\# Prepare next iteration}
		\STATE $\{c^{(k+1)}_{vw}\} \gets \left\{\max\left(c^{(k)}_{vw},s^{(k)}\right),
			\forall (v,w)\in\mathcal{S}_{(k)}\right\}$
		\COMMENT{Update constraints}
		\STATE $\mathcal{S}_{(k+1)} = \left\{\mathcal{S}_{(k)}\cup u\right\}\setminus m^{(k)}$ \COMMENT{Update mergeable nodes}
		\ENDFOR
		\STATE \textbf{return} $\mathcal{L}_{\tau}, \mathcal{L}_{\ell}$
		\ENDFUNCTION
	\end{algorithmic}
\end{algorithm}

\subsection{Reparameterization of the branch length probability $q_\ell$}
\label{app:jacobian}

Because we want to constrain the sum of branch lengths in our probability distribution, we reparameterize $\ell=\left(\ell_i, \ell_j\right)$ as $(s,r) = g(\ell) = \left(\ell_i+\ell_j, \ell_i/(\ell_i+\ell_j)\right)$ and model the distribution of $q_{(s,r)}(s,r) = q_s(s)q_r(r)$, as detailed in~\ref{sec:bayesnj}. In order to recover the probability $q_\ell$ from $q_s(s)q_r(r)$, we must account for this change of variable:
\begin{equation*}
	q_\ell(\ell) = q_{(s,r)}\left(g(\ell)\right)\left|\det J_g(\ell)\right|,
\end{equation*}
where
\begin{equation*}
	J_g(\ell) = \begin{pmatrix}
		1                            & 1                             \\
		\frac{\ell_j}{\ell_i+\ell_j} & -\frac{\ell_i}{\ell_i+\ell_j}
	\end{pmatrix}
\end{equation*}
is the Jacobian matrix of $g$, whose determinant is therefore $-1/(\ell_i+\ell_j) = -1/r$. As a result,
\begin{equation*}
	q_\ell(\ell) = q_s(s)q_r(r)/r.
\end{equation*}

\subsection{Training runs}

\begin{table}[H]
	\rowcolors{1}{white}{black!5}
	\centering
	\resizebox{1\textwidth}{!}{%
		\begin{tabular}{lcccccccccc}
			\toprule
			Model                          & \thead{Starting                                                                                                      \\weights} & \thead{Embedding\\dimensions\\($c_s\vert c_z$)}& \thead{Training\\data} &\thead{Loss\\function} & \thead{Training\\time} & \thead{Batch\\size} & \thead{Scheduled\\epochs} & \thead{Target\\LR} & Warmup & \thead{Selected\\step} \\
			\midrule
			(PF2)                          & Random          & $(128\vert256)$ & BD,LG+G8       & BayesNJ & 62.5h & 16   & 30 & $10^{-4}$       & $10^3$  & 86000 \\
			\textbf{PF2$_\text{MAE}$}      & Random          & $(128\vert256)$ & BD,LG+GC       & MAE     & 26h   & 16   & 30 & $5\cdot10^{-4}$ & $10^3$  & 51000 \\
			\textbf{PF2}                   & (PF2)           & $(128\vert256)$ & BD,LG+G8,multi & BayesNJ & 8.5h  & 1-40 & 30 & $10^{-6}$       & $0.5\%$ & 8000  \\
			\textbf{PF2$_{\text{Cherry}}$} & (PF2)           & $(256\vert512)$ & BD,Cherry      & BayesNJ & 42h   & 6    & 30 & $10^{-5}$       & $0.5\%$ & 41000 \\
			\textbf{PF2$_{\text{SelReg}}$} & (PF2)           & $(256\vert512)$ & BD,SelReg      & BayesNJ & 42h   & 6    & 30 & $10^{-5}$       & $0.5\%$ & 44000 \\
			\textbf{PF2$_{\text{MCMC}}$}   & (PF2)           & $(256\vert512)$ & U+Exp,LG+G8    & BayesNJ & 10h   & 6    & 30 & $10^{-6}$       & $0.5\%$ & 12009 \\
			\bottomrule
		\end{tabular}%
	}
	\caption{Training run parameters. All runs were in a distributed data-parallel setting using 4 H100 GPUs. Final models are shown with their name in bold. More details on the training data in \Cref{sec:data} and \Cref{fig:training}.}
	\label{tab:training}
\end{table}

\begin{figure}[H]
	\centering
	\includegraphics[width=\linewidth]{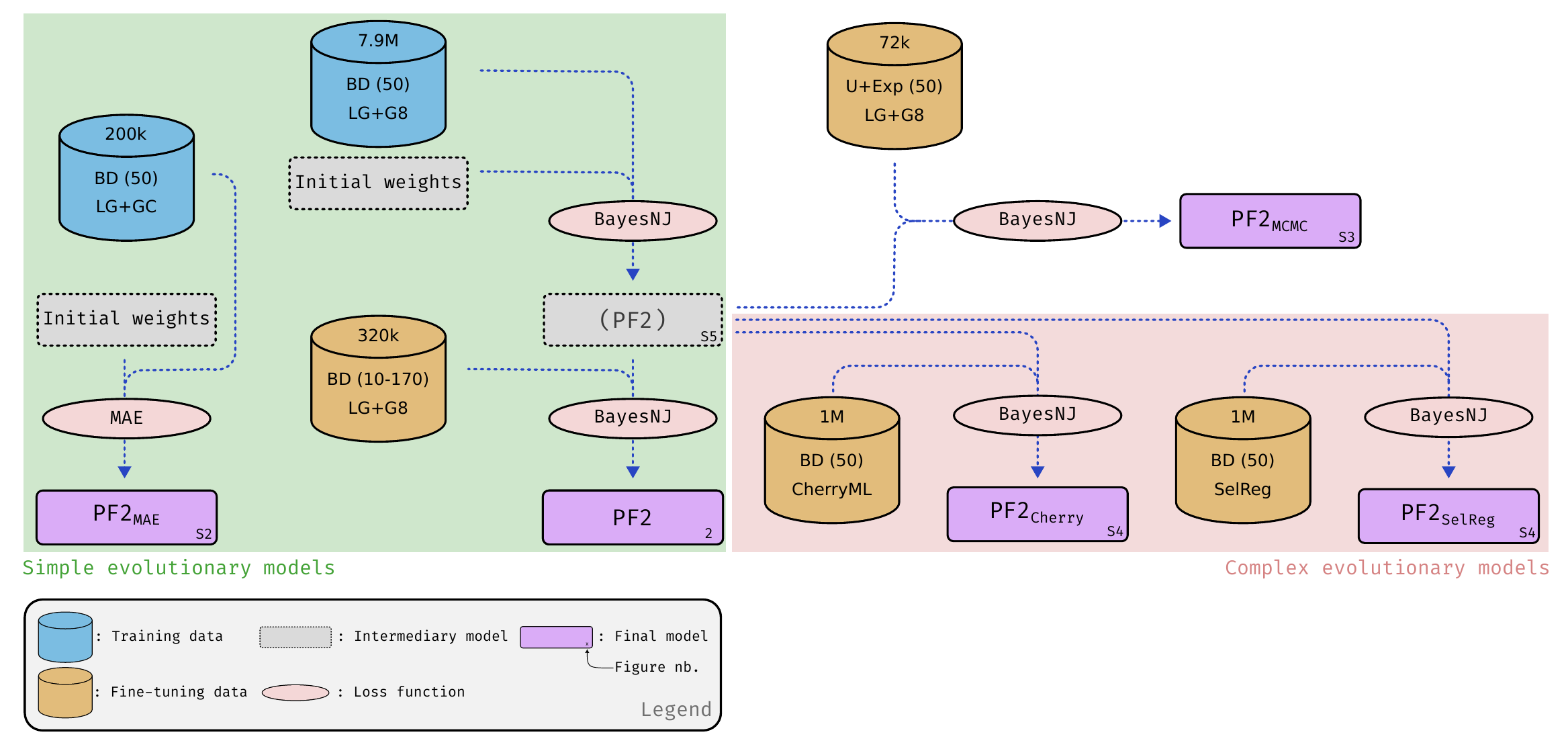}
	\caption{Training setting for all PF2 instances. PF2 instances are shown in rectangles, with a full outline and purple fill if they are final models used for inference in the main results, and a dotted outline and gray fill if they are used as the starting model for fine tuning runs. The loss functions used for each training or fine-tuning runs are shown in ovals. Finally the training (blue) and fine-tuning (yellow) datasets are also shown as cylindrical shapes. For each dataset, the number of training and validation examples is shown in the top section of the cylinder. The simulation priors are shown in the body of each cylinder : (1) the tree topology prior with the training tree size in parentheses, (2) the MSA evolutionary model. The tree priors are either \textit{(BD)}: rescaled Birth-death, described in  \citep{nesterenko2025phyloformer}, or \textit{(U+Exp)}: Uniform tree topology with $\lambda=10$ exponentially distributed branch lengths available in RevBayes \citep{hohna2016RevBayes}. The number of figure in which the performance of a given model is studied is shown in the bottom right corner of the corresponding rectangle. Datasets are described in \Cref{sec:data}}
	\label{fig:training}
\end{figure}

\subsubsection{Training datasets}
\label{sec:data}

\paragraph{From \citep{nesterenko2025phyloformer}:} The 200k LG+GC, Cherry and Selreg datasets, used for training and fine-tuning models were obtained from the original paper. A full description of these datasets is available in the supplementary material of \citep{nesterenko2025phyloformer}

\paragraph{LG+G8 dataset:} This is the main dataset used to train PF2.
the 7,966,499 training trees (and 10,000 validation trees) are simulated with 50 leaves, following the empirically rescaled birth-death procedure described in \citep{nesterenko2025phyloformer}.
MSA were simulated along these trees using IQTree's alisim tool \citep{minh2020IQ-TREE} under the LG evolutionary model. Rate heterogeneity across sites was modeled with an 8 category discrete Gamma. Insertions-deletion events were added using alisim, with identical rates of $2\cdot10^{-3}$, and insertion-deletion length was modeled with the default options: a zipfian distribution with an exponent of 1.7 and a maximum size of 100. These parameters were chosen to yield MSAs with 10\% of gaps on average on trees sampled from our prior distribution, which seems consistent with the gap content of empirical data.

\paragraph{Multi-size LG+G8 dataset:} This dataset was used to fine-tune PF2 to limit the effect of overfitting to the training. Tree-MSA pairs were generated with 10 to 170 leaves (with a step-size of 10) using the same procedure as the main LG+G8 dataset described above. We simulated 20,000 training examples and 1,000 validation examples for each tree size.

\paragraph{MCMC fine-tuning dataset:} This dataset was used to fine tune PF2 under the prior used in RevBayes \citep{hohna2016RevBayes} to estimate the posterior distribution shown in \Cref{fig:calibration}. 72,007 training (resp. 1000 validation) trees were simulated with RevBayes with uniformly distributed topologies and branch lengths sampled from an exponential distribution with $\lambda=10$. MSAs were simulated along those trees using the same procedure as the main LG+G8 and the multi-size LG+G8 datasets described above.

\paragraph{Potts Model training dataset:} \citet{pagnaniGenerativeContinuousTime2025a} introduced a continuous-time sequence evolution model parametrized by a given Potts model. This can be used to simulate realistic, family-specific MSAs along a phylogenetic tree that accurately reflect real epistatic effects. We used this method to generate 2 million (tree,MSA) pairs under a Potts model fit using \texttt{adabmDCA} \cite{rossetAdabmDCA20AFlexible2026} on the response regulator PF00072 PFAM family \cite{mistryPfamProteinFamilies2021a}. 5,000 of these were used as a separate testing set (see. \Cref{fig:pastek-cherry}), 5,000 as validation examples and 1.5 million as training examples. This was used to train a PF2 instance from scratch.
\label{sec:potts}

\subsection{Fine tuning on different tree-sizes}
\label{sec:fine-tuning-batch-size}
In our current implementation, the BayesNJ module only accepts batches with MSA with the same number of taxa. In order to avoid wasting GPU memory we use a dynamic batch-size during fine-tuning, where the number of examples for a batch with MSAs of a certain size is set to use as much GPU memory as it can. Since batch-size and learning rate are tightly linked, we define the learning-rate schedule for MSAs with 50 sequences and rescale the learning rate applied at a certain batch by the ratio of the current batch-size and the one for MSAs with 50 sequences. We also set a maximum batch-size equal to the one chosen for MSAs of 50 sequences, this ensures that the model sees enough batches for small MSAs during fine-tuning.

\subsection{The Cherry dataset in~\citet{nesterenko2025phyloformer} contains unrealistic amounts of coevolution between sites}
\label{app:cherry}
The Cherry dataset simulated in ~\citet{nesterenko2025phyloformer} was simulated by using an incorrect rate matrix.
As a result, the amount of co-evolution among pairs of coevolving residues was superior to what can be found in empirical data.
The cause of this high amount of coevolution is a mistake in the use of the Cherry matrix \cite{prillo2023cherryml}.
Instead of using the Cherry rate matrix to model pairs of interacting sites, the authors used the product of the rate matrix with its stationary frequencies.
In standard models of molecular evolution, it is customary to represent a reversible substitution rate matrix $Q$ as a product of an exchangeability matrix $R$ and stationary frequencies $F$: $Q = R\times F$.
The Cherry dataset in ~\cite{nesterenko2025phyloformer} was actually simulated according to a matrix $Q' = R \times F \times F = Q \times F $.
The resulting data provides an example of data with extreme amounts of coevolution, which we use as an example of strong departure from standard models of sequence evolution as implemented in e.g., IQ-Tree \cite{minh2020IQ-TREE}.

\subsection{Break even analysis}
\label{sec:break-even}
We analyze the total compute time needed to obtain inferred trees for PF2 and ML methods in \Cref{tab:break-even}. While PF2's training time on 7.9M (tree,alignment) pairs was 62h on GPUs, the simulation of the 7.9M pairs took approximately 1450h of CPU time.
Similarly, PF2's fine-tuning time on $16\times20$k (tree,alignment) pairs was 8h30 on GPUs while the simulation of these pairs took 288 CPU hours.
This is because we discard all alignments that contain identical sequences, which forces us to simulate many more pairs ($6\times$ more for MSAs with 50 sequences, and up to $60\times$ more for alignments with 170 tips) than we actually use.
Other simulation protocols may be much shorter, and we plan to improve this step of our pipeline. In the following, we report break-even points both with and without the simulation. Because of this cost, if we only infer trees for MSAs of 50 tips, PF2 would break even in terms of compute time after 230k tree inferences with IQTree with the LG+GC model (9042 tree inferences without simulation), 16k with IQTree and ModelFinder (629), or 4.3M FastTree inferences (169047). If we focus on trees of 100 tips this goes down to 91k IQTree with LG+GC (3577), 8k IQTree with ModelFinder (315) and 2.2M FastTree inferences (86489).
While the break-even points are quite high for point estimation, this changes radically when sampling a large number of trees from the learned posterior distribution. Since PF2 can sample from the posterior in parallel using GPUs it is very fast, and for trees with 50 tips we reach the break even point w.r.t RevBayes in only 11 sampling runs of 10k iterations. In practice it might be even lower because some of the MCMC samples must be discarded during the burn-in phase.

\begin{table}[H]
	\centering
	\begin{tabular}{rcccccc}
		\toprule
		             & Simulation & Training & MAP ($n=50$) & MAP ($n=100$) & 10k Samples ($n=50$) \\\midrule
		\textbf{PF2} & 1735h      & 71h      & 0.6          & 1.4           & 38.8                 \\
		IQTree       & -          & -        & 28.9         & 73.1          & -                    \\
		IQTree+MF    & -          & -        & 405.5        & 826.6         & -                    \\
		FastTree     & -          & -        & 2.1          & 4.4           & -                    \\
		RevBayes     & -          & -        & -            & -             & 172h                 \\ 
		\bottomrule
	\end{tabular}%
	\caption{Time per inference and upfront tasks for PF2 vs ML and MCMC methods. Simulation times are estimated from Snakemake pipeline reports and are reported in CPU time. In practice this is highly parallelisable so wall time would be much shorter. Here we compare PF2 and other methods on 2 tasks. (1) Point estimations (for 50-sequence and 100-sequence MSAs) where we compare ourselves to maximum-likelihood methods: IQTree, IQTree with ModelFinder and FastTree, and (2) Sampling 10k trees from the posterior for a 50-sequence MSA, where we compare ourselves to RevBayes.
		Times for maximum likelihood methods are obtained from \citep{nesterenko2025phyloformer}.
		The mean time to sample 10k trees from the posterior distribution, replicated over 50 50-sequence MSAs, using PF2 is reported. PF2 samples from the posterior in batches of 500 parallel samples.
		For the RevBayes estimate, we measured how many trees were able to be sampled during a fixed 20h duration and extrapolate.
		Simulation and Training time for PF2 are the sum of the times for the main training and fine-tuning phases.
		Reported times are in seconds unless specified otherwise.}
	\label{tab:break-even}
\end{table}

\subsection{Empirical data experiments}
\label{app:empirical}

Data sets from \citet{zhouEvaluatingFastMaximum2018}, filtered to retain only alignments best fitted by the LG+Gamma model of sequence evolution as in \citet{nesterenko2025phyloformer} were realigned with Prank \citep{loytynojaPhylogenyawareAlignmentPRANK2014}. Prank is a highly accurate alignment algorithm that ensures that gaps are positioned in a way that is consistent with a phylogeny that it estimates itself, using a distance method from a fast alignment obtained with Mafft \citep{katoh_multiple_2013}.

As discussed in the main manuscript, the distribution of Robinson-Foulds errors is slightly larger than for other methods on NagyA1 and ShenA9. We reasoned that this might by caused by a discrepancy between the training and test data. Specifically, PF2 was trained on alignments containing gaps that were all consistent with the phylogeny: because we simulate the sequences with a Markovian process, we know precisely which positions align together and where to include gaps in case of insertions-deletions. By contrast on empirical data, we must rely on multiple alignment algorithms that can make mistakes, including in their choice of gap placement. Because other reconstruction methods in our benchmarks treat gaps as missing data, erroneous gaps only hurts them by discarding signal. PF2 not only loses the same signal but was trained to extract information from these gaps which can lead to more error. 

To test this hypothesis we compared phylogenetic reconstruction accuracy when alignments are built with Prank \citep{loytynojaPhylogenyawareAlignmentPRANK2014} provided with the reference species tree, leading to a much more accurate alignment---and in particular gap placements. This experiment (Figure~\ref{fig:prank_effect}, boxplots Prank+guide) obviously doesn't estimate how well the benchmarked methods are able to reconstruct phylogenies---because in practice we wouldn't know the species tree and could not obtain such a high-quality alignment---but indicates how sensitive these methods are to the alignment quality. As expected, all methods improve under this setting, but PF2 becomes the most accurate. This is consistent with our hypothesis: improving the alignment not only fixes a source of error---incorrect gaps---that affected PF2 more than other methods, but it also provides a signal that only PF2 can use. Although it was trained on data similar to PF2, PF1 does not seem as capable of using gap information, which may be lost during the construction of a pairwise distance matrix, whereas PF2 can propagate gap information from tips to root thanks to BayesNJ.

Figure~\ref{fig:prank_effect} also shows performances on the original, lower quality alignments~\cite{wong2008alignment, loytynoja2008phylogeny} provided by~\citet{zhouEvaluatingFastMaximum2018}. Consistently, these alignments with a larger number of gaps not consistent with the phylogeny dramatically decrease the performance of PF2 compared to other methods---which just ignore these gaps whereas PF2 treat them as signal.
As further evidence of PF2's use of gap information for phylogenetic reconstruction, we measured on the PRANK alignments the consistency of gaps with the phylogeny using the consistency index \citep{felsenstein2003}.
A character is consistent with a tree topology if it only appeared once on the tree.
Characters may appear several times due to convergent evolution, but this is expected to be rare.
We found that 139,496 gap sites were more consistent on PF2 trees whereas only 21,959 sites were more consistent on IQTREE trees (with 350,391 gap sites tied). This confirms that gap information contributed to phylogenetic reconstruction with PF2, which therefore reconstructed trees more consistent with the gaps.

Furthermore, we show in~\Cref{fig:prank_ppc} that a posterior predictive check procedure can detect when PF2 behaves poorly at the dataset level. The procedure measures the ability to sample alignments similar to the input from trees generated from the estimated posterior distribution. It clearly highlights a failure on the two problematic datasets NagyA1 and ShenA9.

\begin{figure}
    \centering
    \includegraphics[width=0.95\linewidth]{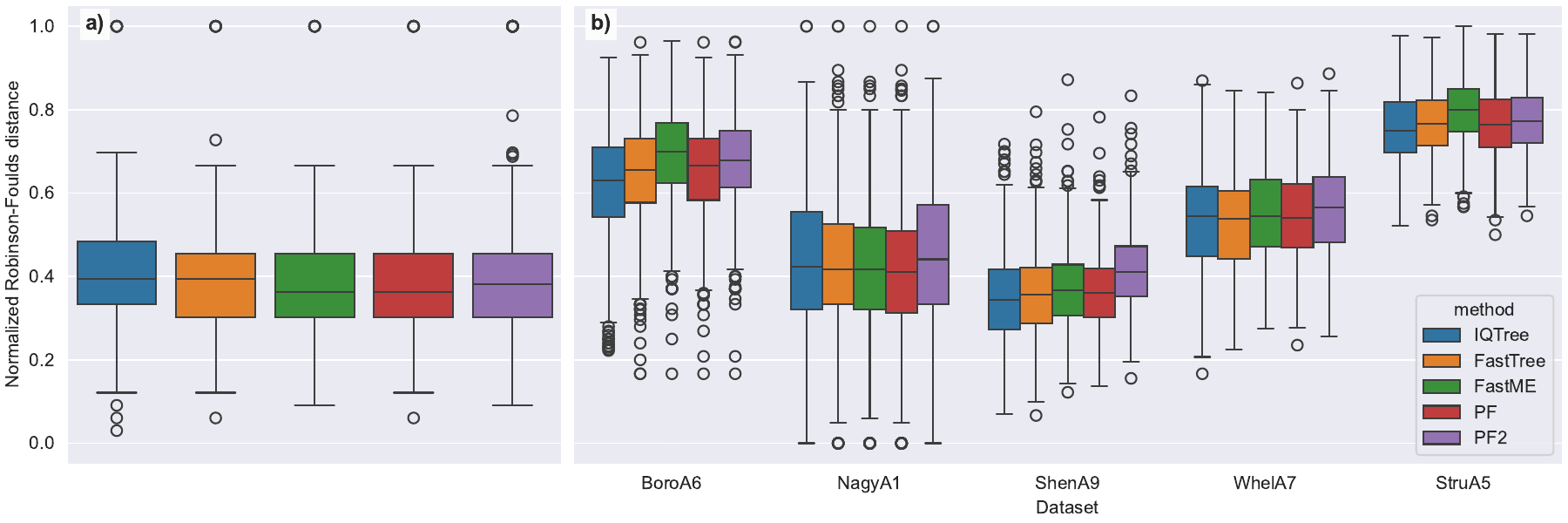}
    \caption{Topological reconstruction accuracy on empirical datasets. The topological accuracy is measured as the normalized Robinson-Foulds distance between the inferred gene tree, and a reference species tree inferred on the concatenation of all gene MSAs using maximum-likelihood estimation.
    This experiment is done on a cyanobacteria dataset introduced in \citet{szollosiDataEfficientExploration2013} (left), 
    and a set of diverse protein datasets introduced in \citet{zhouEvaluatingFastMaximum2018} (right) and realigned with PRANK \cite{loytynojaPhylogenyawareAlignmentPRANK2014}.
    }
    \label{fig:cyano_rokas}
\end{figure}

\begin{figure}
    \centering
    \includegraphics[width=0.8\linewidth]{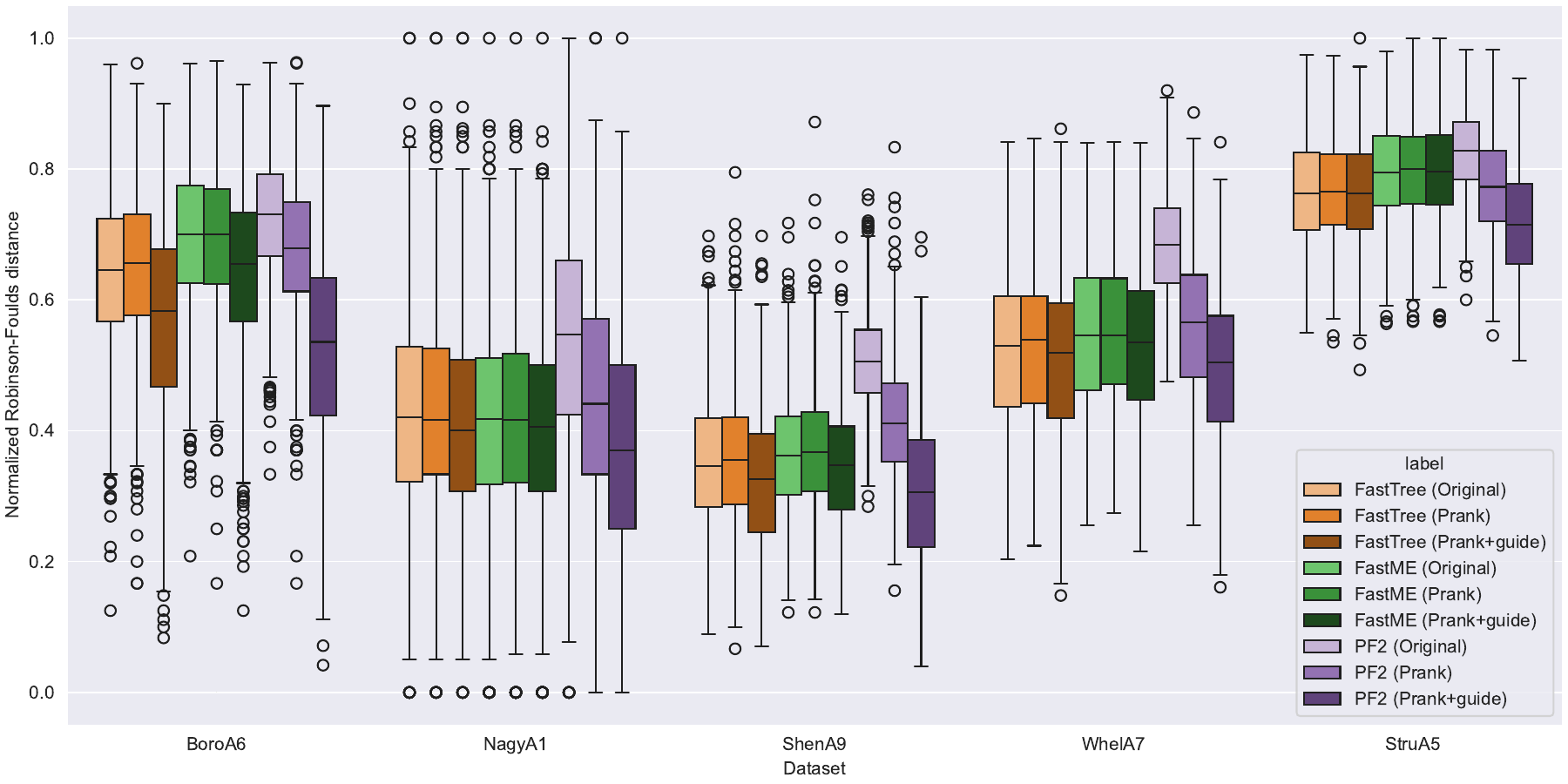}
    \caption{Topological reconstruction accuracy on a set of diverse protein datasets introduced in~\citet{zhouEvaluatingFastMaximum2018}, with three different alignment procedures. The topological accuracy is measured as the normalized Robinson-Foulds distance between the inferred gene tree, and a reference species tree inferred on the concatenation of all gene MSAs using maximum-likelihood estimation.
Alignments are done under the same original procedure as~\citet{zhouEvaluatingFastMaximum2018} (leftmost, clear color), with PRANK~\cite{loytynojaPhylogenyawareAlignmentPRANK2014} (middle, intermediate color) and with PRANK using the concatenate species tree as a guide (right, darkest colors).}
    \label{fig:prank_effect}
\end{figure}

\begin{figure}
    \centering
    \includegraphics[width=0.9\linewidth]{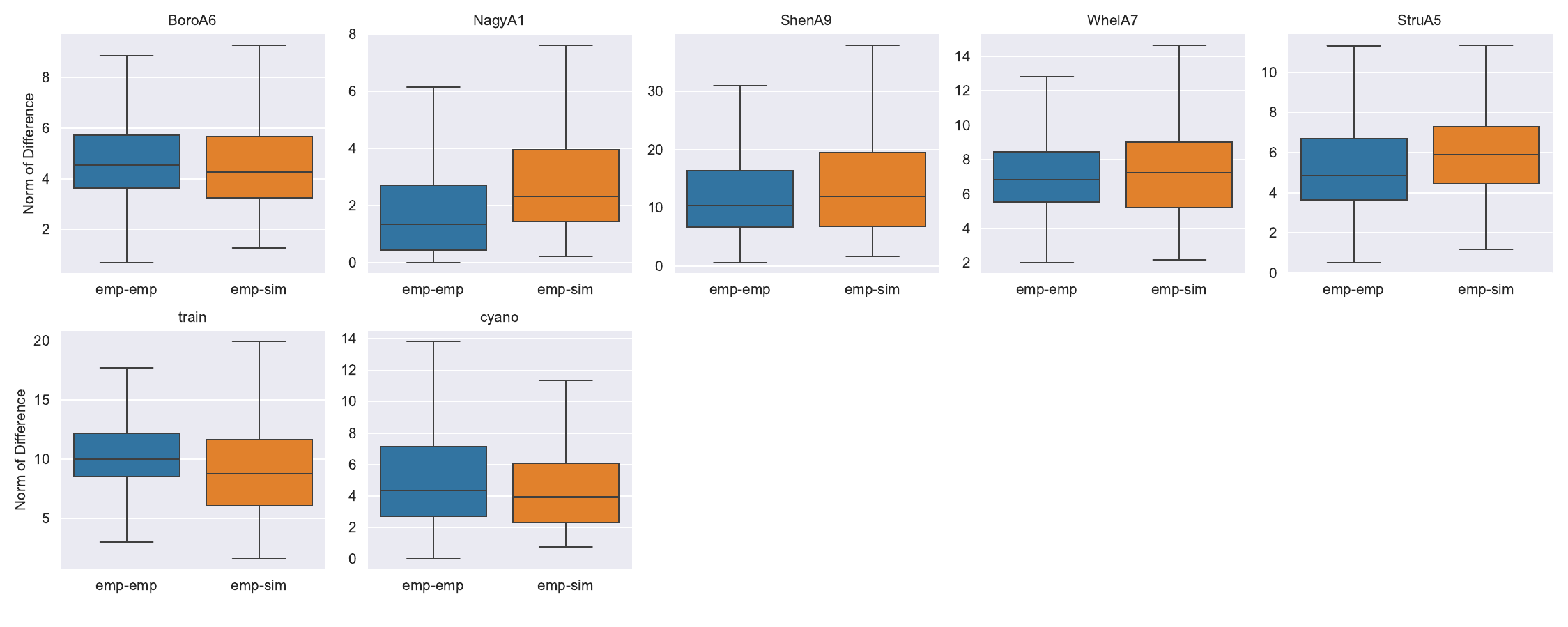}
    \caption{Posterior predictive checks on empirical data. For each tree estimate on an empirical MSA ($MSA_{emp,i}$), a synthetic MSA ($MSA_{sim,i}$) is simulated along the estimated tree using our standard approach. Both MSA types are encoded as pairwise Hamming distance matrices (ignoring gaps).
    The distribution of the Frobenius norm of the $MSA_{emp,i}-MSA_{sim,i}$ difference for each empirical MSA is shown in the orange boxplots for each dataset in \Cref{fig:cyano_rokas}. 
    This distribution can be compared to the distribution between pairs of empirical MSAs: $MSA_{emp,i}-MSA_{emp,j}$, shown in the blue boxplots. 
    Results for training data are also shown in the lower-left facet of the plot.}
    \label{fig:prank_ppc}
\end{figure}



\newpage
\subsection{Supplementary Figures}

\begin{figure}[H]
	\centering
    \includegraphics[width=\linewidth]{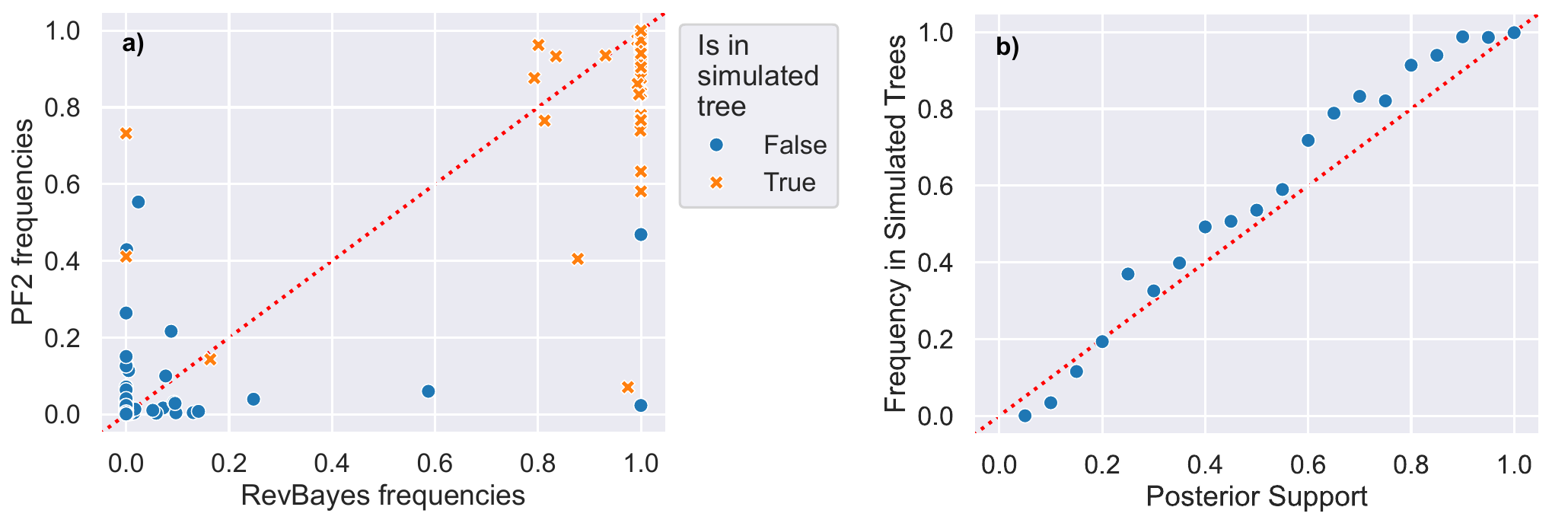}
	\caption{\textbf{a)}: Comparison of split frequencies over samples from the posterior of a single 50 sequences MSA, between RevBayes MCMC (x-axis) and PF2 (y-axis). The orange cross marker indicates splits that are present in the tree along which the MSA used for sampling has been simulated. \textbf{b)}: simulation-based calibration comparing branch supports, \emph{i.e.}, their frequency in samples from the posterior estimate given by a trained PF2 (x-axis) to the frequency with which splits with a given PF2 support are true, i.e. found in the tree used for simulation (y-axis)
		Support values are binned by steps of $0.05$.}
	\label{fig:calibration}
\end{figure}

\begin{figure}[H]
	\centering
	\includegraphics[width=\linewidth]{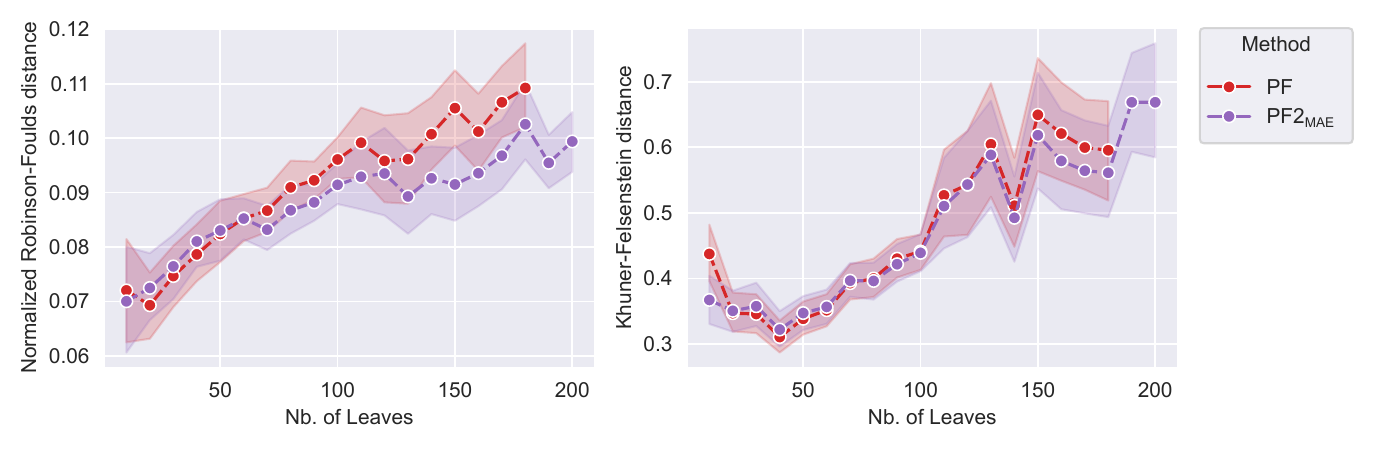}
	\caption{How does PF2 trained with an L1 loss compare to PF? PF2 was trained under the same conditions as the original PF on LGGC trees with an L1 loss on pairwise phylogenetic distances. The Robinson-Foulds distance (left) shows the topological reconstruction accuracy, while the Kuhner-Felsenstein distance (right) takes both topology and branch-lengths into account.}
	\label{fig:ablation}
\end{figure}

\begin{figure}[H]
	\centering
	\includegraphics[width=0.7\linewidth]{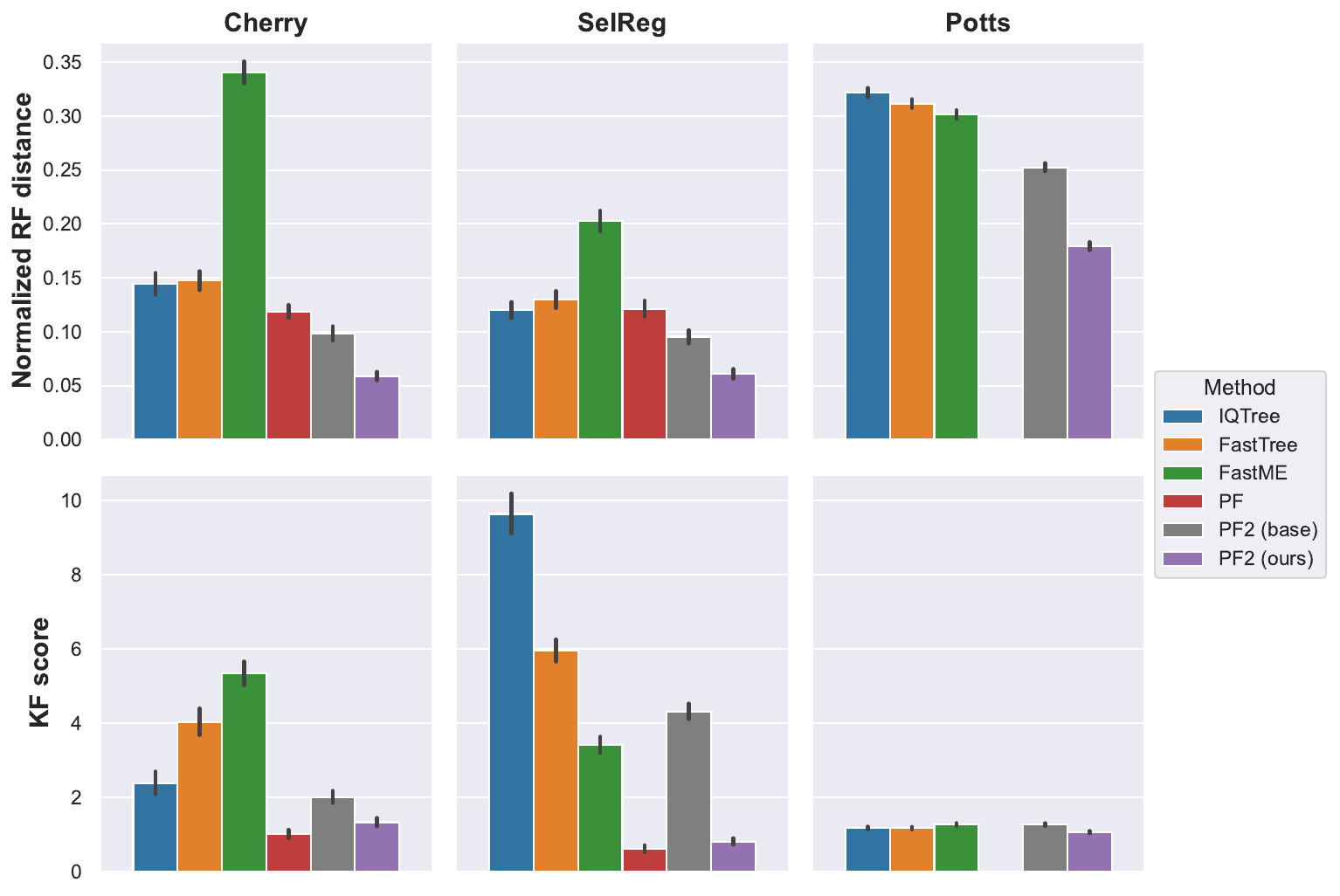}
	\caption{ Average PF2 performance under intractable-likelihood models, measured on 50-tip trees. 
    The tree reconstruction performance was measured under 3 different sequence-evolution models for which the likelihood-computation is hard or intractable. Each sequence model is shown in a column, from left to right: Cherry \cite{prillo2023cherryml} which simulates pairs of co-evolving sites, SelReg \citep{duchemin2022evaluation} which simulates sites under different selective scenarii, and a Potts model fit on the PFAM PF00072 family (also see \ref{sec:potts}).
    PF (red) and PF2 (purple) represent instances of the model that have been fine-tuned on data simulated under the corresponding sequence evolution model. Since the Potts model was not used in \cite{nesterenko2025phyloformer}, PF performance measures do not exist for this dataset. For each dataset the error bar shows the 95\% CI computed with 1000 bootstrap samples.%
    }
	\label{fig:pastek-cherry}
\end{figure}

\begin{figure}[H]
	\centering
	\includegraphics[width=0.7\linewidth]{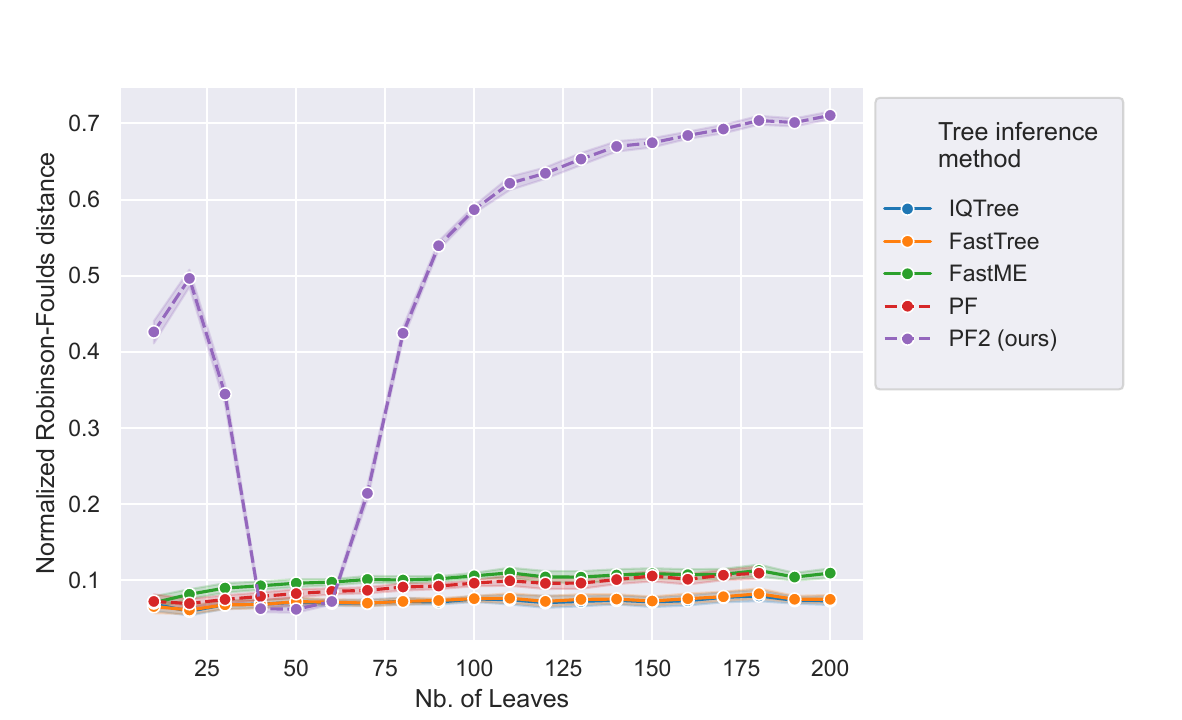}
	\caption{Topological performance for non fine-tuned topology-only Phyloformer~2, measured by the normalized Robinson-Foulds distance. The MSA dataset and compared method results are the same as \Cref{fig:PF2-LGGC}}
	\label{fig:PF2_no_FT}
\end{figure}

\begin{figure}[H]
	\centering
	\includegraphics[width=0.7\linewidth]{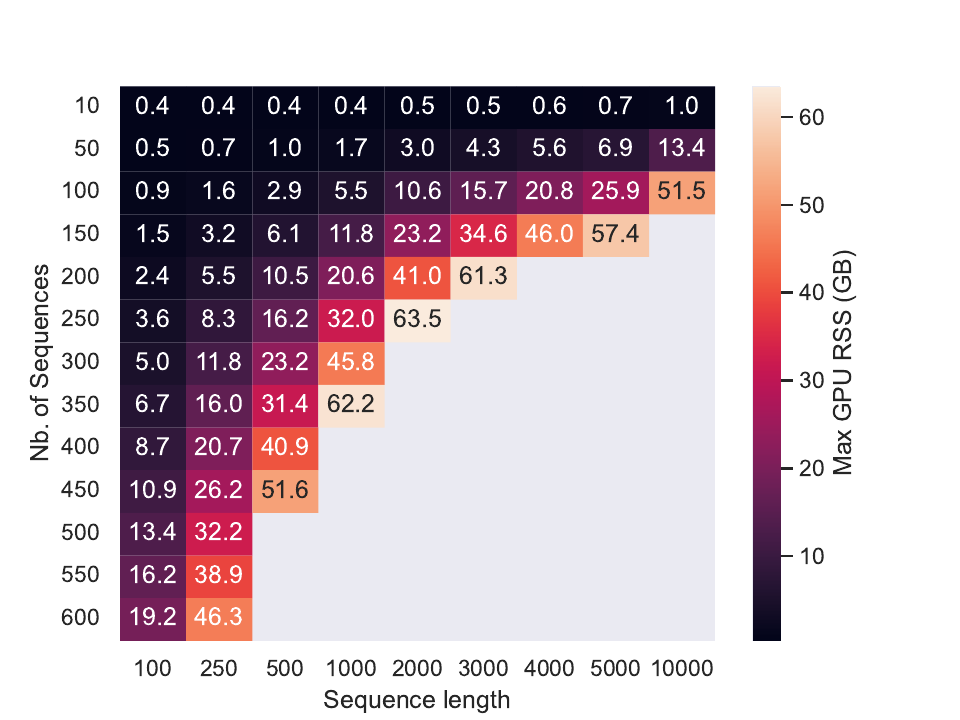}
	\caption{PF2 memory usage (in GB) scaling w.r.t number of sequences and sequence length measured on an H100 GPU}
	\label{fig:mem_stress_test}
\end{figure}

\begin{figure}[H]
	\centering
	\includegraphics[width=0.9\linewidth]{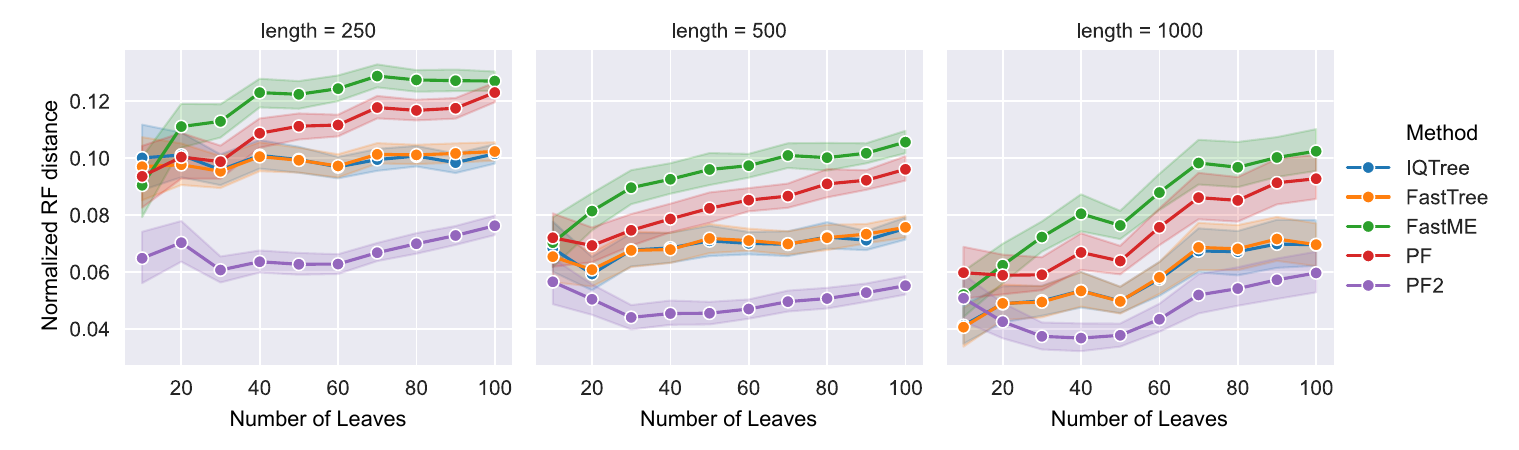}
	\caption{Topological reconstruction accuracy for PF2 and other methods for different sequence lengths. MSAs of different lengths are simulated on the same set of phylogenetic trees for the 3 panels.}
	\label{fig:MSA_length_overfiting}
\end{figure}
\end{document}